\documentclass[aps, prl, reprint, superscriptaddress, amsmath, amssymb]{revtex4-2}

\usepackage{braket}
\usepackage{graphicx}% Include figure files
\usepackage{dcolumn}% Align table columns on decimal point
\usepackage{bm}% bold math
\usepackage{color}
\usepackage{soul} % overstriking \st{}
\usepackage[colorlinks, citecolor=blue,urlcolor=blue,linkcolor=blue]{hyperref}% add hypertext capabilities
\begin{document}

\title{Imaginary-Stark Skin Effect}

\author{Heng Lin}
    \affiliation{State Key Laboratory of Low-Dimensional Quantum Physics and Department of Physics, Tsinghua University, Beijing 100084, China}
\author{Jinghui Pi}
    \email{pijh14@gmail.com}
    \affiliation{State Key Laboratory of Low-Dimensional Quantum Physics and Department of Physics, Tsinghua University, Beijing 100084, China}
\author{Yunyao Qi}
    \affiliation{State Key Laboratory of Low-Dimensional Quantum Physics and Department of Physics, Tsinghua University, Beijing 100084, China}
\author{Wei Qin}   
    \affiliation{Center for Joint Quantum Studies and Department of Physics, Tianjin University, Tianjin 300350, China}
\author{Franco Nori}
    \affiliation{Center for Quantum Computing, RIKEN, Wako-shi, Saitama 351-0198, Japan}
    \affiliation{Department of Physics, The University of Michigan,
	Ann Arbor, Michigan 48109-1040, USA}
\author{Gui-Lu Long}%
    \email{gllong@tsinghua.edu.cn}
    \affiliation{State Key Laboratory of Low-Dimensional Quantum Physics and Department of Physics, Tsinghua University, Beijing 100084, China}
    \affiliation{Frontier Science Center for Quantum Information, Beijing 100084, China}
    \affiliation{Beijing Academy of Quantum Information Sciences, Beijing 100193, China}

\begin{abstract}
A unique phenomenon in non-Hermitian systems is the non-Hermitian skin effect (NHSE), namely the boundary localization of continuous-spectrum eigenstates. However, studies on the NHSE in systems without translational invariance are still limited. Here, we unveil a new class of NHSE, dubbed the imaginary-Stark skin effect (ISSE), in a one-dimensional lossy lattice with a spatially increasing loss rate. {\it This ISSE is beyond the framework of non-Bloch band theory and exhibits intriguing properties significantly different from the conventional NHSE.} 
Specifically, the energy spectrum of our model has a T-shaped feature, with approximately half of the eigenstates localized at the left boundary. Furthermore, each skin mode can be expressed as a single stable, exponentially-decaying wave within the bulk region. 
Such peculiar behaviors are analyzed via the transfer-matrix method, whose eigendecomposition quantifies the formation of the ISSE.
Our work provides new insights into the NHSE in systems without translational symmetry and contributes to the understanding of non-Hermitian systems.
\end{abstract}

\maketitle

In closed systems, the Hermiticity of Hamiltonians is a fundamental postulate. In contrast, for open systems, non-Hermiticity emerges and can be described by effective Hamiltonians \cite{El2018Non}, displaying peculiar properties and potential applications in various fields \cite{PhysRevLett.77.570, PhysRevB.56.8651, PhysRevB.58.8384, PhysRevLett.80.5243, berry2004physics, PhysRevLett.102.065703, PhysRevLett.103.093902, Ruter2010Observation, Feng2011Nonreciprocal,  Regensburger2012Parity-time, Zhen2015Spawning,  Poli2015Selective, longhi2015robust, Chen2017Exceptional, Xiao2017Observation, Bahari2017Nonreciprocal, Bandres2018Topological, Bliokh2019Topological, Özdemir2019Parity–time, ashida2020non, Leefmans2022Topological, Arkhipov2023Dynamically, PhysRevResearch.5.L032026}. A unique feature of non-Hermitian systems is the non-Hermitian skin effect (NHSE), namely the boundary localization of bulk band eigenstates, causing a high sensitivity of the spectrum to the boundary conditions \cite{PhysRevB.97.121401, prl_wang, PhysRevLett.121.136802}. The NHSE can induce novel phenomena without Hermitian counterparts, including unidirectional physical effects \cite{Song_2019, Wanjura_2020, Xue_2022}, critical phenomena \cite{Li_2020, Liu_2020Helical, Yokomizo_2021, PhysRevLett.127.116801}, and geometry-related effects in high dimensions \cite{Sun_2021, Zhang_2022universal, Li_2022gain, Zhu_2022, Wu_2022complex}. 
Notably, the NHSE has also been observed across diverse experimental platforms, such as active mechanical materials \cite{brandenbourger2019non,palacios2021guided,PhysRevLett.131.207201}, electrical circuits \cite{helbig2020generalized,zhang2021observation,10.21468/SciPostPhys.16.1.002}, optical systems \cite{doi:10.1126/science.aaz8727, xiao2020non, Liu2022Complex, Leefmans2024Topological}, and cold atoms \cite{PhysRevLett.124.070402}.

Thus far, most studies of the NHSE focus on translation-invariant systems, amenable to non-Bloch band theory \cite{prl_wang,PhysRevLett.123.066404,PhysRevLett.125.126402, PhysRevX.14.021011}. Specifically, the extended Bloch waves are replaced by exponentially-decaying waves, drastically changing the system topology and thus reshaping the conventional bulk-boundary correspondence \cite{PhysRevLett.118.040401, PhysRevLett.120.146402, PhysRevX.8.031079, kawabata2019topological, PhysRevX.9.041015, PhysRevB.99.235112, PhysRevLett.124.056802, PhysRevLett.123.066405, Xiong_2018, PhysRevB.99.201103, PhysRevLett.122.076801, PhysRevLett.123.016805, PhysRevB.102.235151}. 
Recently, studies on the NHSE have also been extended to translational symmetry-breaking systems, mainly including quasicrystals \cite{PhysRevB.100.054301, PhysRevB.103.014201, PhysRevB.103.014203, PhysRevB.106.024202, Manna2023Inner, PhysRevB.107.064305, PhysRevB.108.014202}, and systems with disorder \cite{PhysRevB.103.L140201, PhysRevB.103.144202, PhysRevB.104.L121101, PhysRevB.104.L241402, PhysRevB.106.014207, PhysRevB.106.064208, PhysRevB.107.144204, Jin2023Reentrant} or single impurity \cite{Li2021Impurity, PhysRevLett.127.116801, PhysRevResearch.5.033058}. In a system with a single impurity, the translational symmetry is preserved outside the impurity, and thus, the exponentially-decaying wave ansatz of non-Bloch theory still holds. For quasicrystals and disordered systems, the NHSE is typically quantified by the inverse participation ratio, which offers only a superficial description of the NHSE and fails to analyze the wave function comprehensively. Therefore, it remains unclear whether the skin modes in transnational symmetry-breaking systems can exhibit distinct properties, compared to those in translation-invariant systems. Furthermore, the accurate quantification of the NHSE 
is also unsolved when non-Bloch band theory breaks down.

\begin{figure}[t]
    \includegraphics[width=8.8cm]{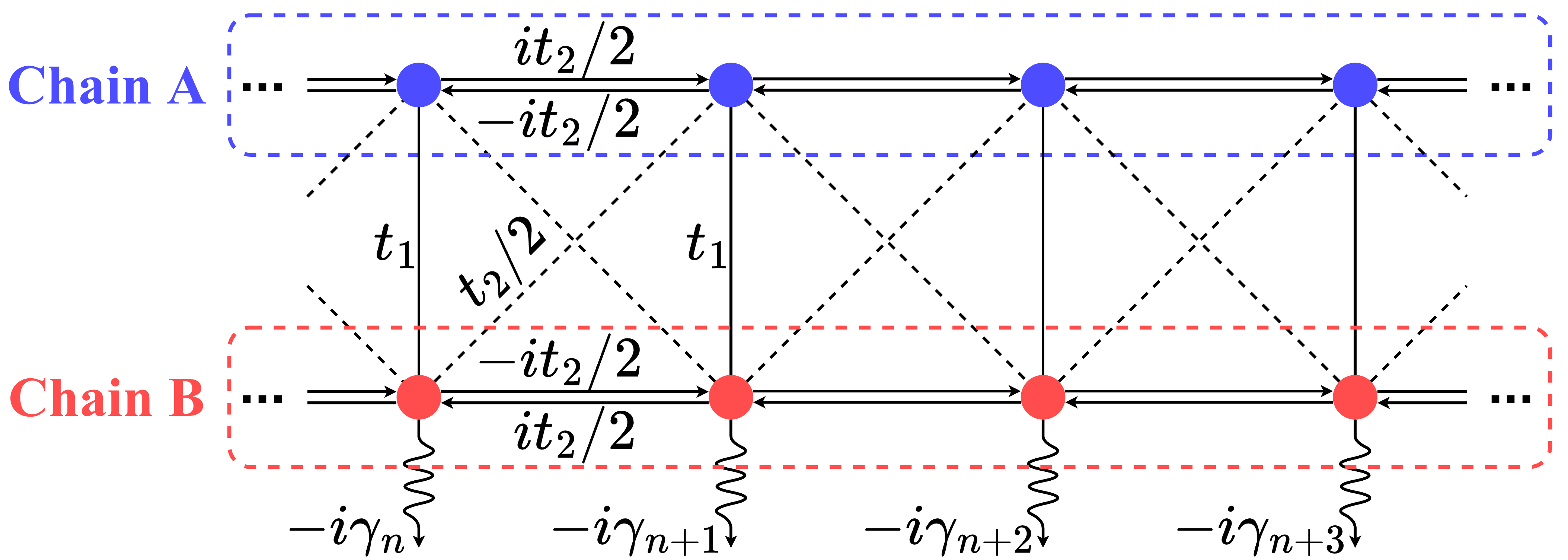}
    \caption{\label{fig:model} A schematic diagram of the non-Hermitian lattice. 
    %\textcolor{red}{$t_1$ and $t_2$ are the hopping strength of Hermitian part.}
    %The Hermitian part is given by the hopping parameters $t_1$ and $t_2$. The non-Hermitian part is a monotonically increasing dissipation $-i\gamma_n$ in chain B, which can be viewed as an imaginary electric potential. 
    The Hermitian interaction is characterized by the hopping parameters $t_1$ and $t_2$, while the monotonically increasing dissipation $\gamma_n$ features  the non-Hermitian processes.
    Here, $n$ labels the lattice index.
    %A one-dimensional non-Hermitian lattice with a monotonically increasing \textcolor{red}{dissipation $-i\gamma_n$}, which can be viewed as an imaginary electric potential.
    }
\end{figure}

\begin{figure*}[t]
\includegraphics[width=17.2cm]{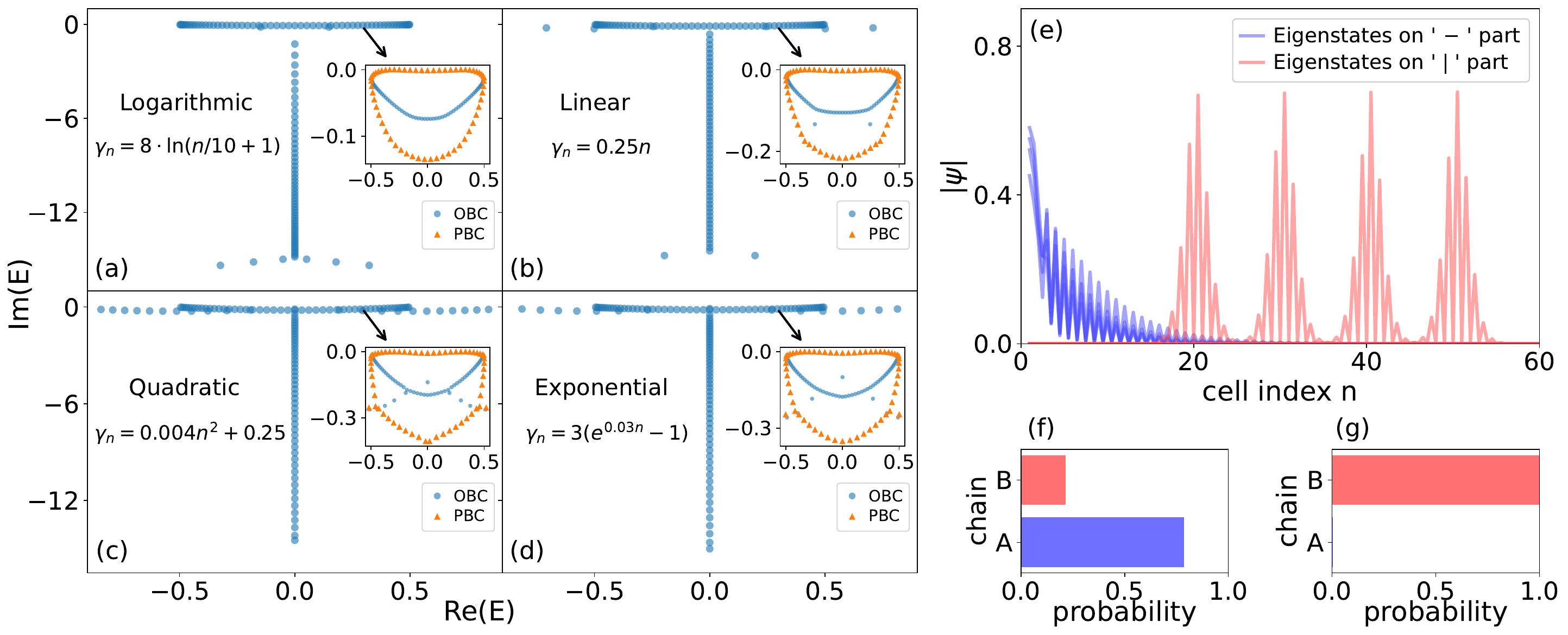}
    \caption{\label{fig:phenomenon} (a)-(d) Energy spectra for various loss rate functions $\gamma_n$. Insets refer to the `$-$' horizontal parts of the T-shaped spectra. (e) Profile of the eigenstates from the `$-$' and `$|$' parts of the T-shaped spectrum.  Eight eigenstates are displayed, of which four are in the `$-$' part, and the other four are in  the `$|$' vertical part. (f), (g) Average probabilities $\Sigma_{n}{|\psi_n^S|^2}$ of a particle in chains A and B, where $S \in \{A, B\}$, and this summation is averaged over eigenstates in either the `$-$' (f) or `$|$' (g) part of the T-shaped spectrum. In (e)-(g), $\gamma_n = 0.25 n$, and in (a)-(g), $t_1=0.4$, $t_2 = 0.5$, and the chain length $L = 60$.}
\end{figure*}

Here, we present a new class of NHSE, i.e., the imaginary-Stark skin effect (ISSE), which is beyond the framework of non-Bloch band theory. {\it The ISSE demonstrates a distinct behavior %from 
compared to the NHSE in translation-invariant systems, and can be studied by the asymptotic and convergence properties of the transfer-matrix eigenvalues. 
This fills an existing gap in the research %of
on NHSE in translational symmetry-breaking systems.} We consider a 1D lossy lattice with a unidirectionally increasing loss rate. This scenario resembles a lattice subjected to a leftward imaginary field, and the energy spectrum of such a system displays a T-shaped feature \cite{PhysRevB.107.L140302}, with its upper-half eigenstates localized at the left boundary. Surprisingly, these skin modes exhibit an almost uniform decay rate within the bulk region, despite the broken translational invariance. Moreover, our numerical results indicate that these skin modes can be approximately expressed as a {\it single} wave within the bulk region. Therefore, the ISSE fundamentally differs from the conventional NHSE described by non-Bloch band theory (e.g., bipolar \cite{PhysRevLett.123.246801}, corner \cite{PhysRevLett.123.016805,PhysRevB.102.205118}, and geometry-dependent skin effect \cite{Zhang_2022universal}), where each skin mode comprises {\it two} exponentially-decaying waves with an identical decay rate. We employ the transfer-matrix method to establish a connection between the ISSE and the convergence rate of the eigenvalues of the transfer-matrix. The wave function is divided into two parts based on the eigendecomposition of the transfer-matrix, with one part dominating the behavior of the skin modes in the bulk, accounting for the peculiar behavior of the ISSE.

{\it Model.---} 
We consider a single particle in a one-dimensional lossy lattice (see Fig.~\ref{fig:model}), which can be divided into two chains, labeled A and B. 
We define $\ket{n, A(B)}$ as a basis state, where the particle is in the $n$-th site of chain A(B). The projection of the wave function $\ket{\psi}$ is denoted as $\psi_n^{A(B)} = \braket{n, A(B) | \psi}$ and $\psi = (\psi_1^A,\  \psi_1^B,\ \psi_2^A,\ \psi_2^B, ...)$.
Then, the eigenequations of this model can be expressed as
\begin{align}\label{eqn:hamiltonian}
    E \psi_{n}^{A} = &\;  t_1 \psi_{n}^{B} + \frac{t_2}{2} (\psi_{n-1}^{B} + \psi_{n+1}^{B}) + i\frac{t_2}{2} (\psi_{n-1}^{A} - \psi_{n+1}^{A}), \nonumber\\
    E \psi_{n}^{B} = &\;  t_1 \psi_{n}^{A} + \frac{t_2}{2} (\psi_{n-1}^{A} + \psi_{n+1}^{A}) - i\frac{t_2}{2} (\psi_{n-1}^{B} - \psi_{n+1}^{B}) \nonumber\\  & -i\gamma_n \psi_{n}^{B},
\end{align}
where $t_1$ and $t_2$ are the hopping parameters, and $\gamma_n$ represents the loss rate %in 
at the $n$-th site of chain B. When $\gamma_n$ is uniform, this model can be transformed into the non-Hermitian Su-Schrieffer-Heeger model with asymmetric hopping, whose NHSE has been studied \cite{prl_wang}. This manuscript focuses on the case where $\gamma_n$ monotonically increases and diverges as $n$ approaches infinity. The energy spectrum of such a model demonstrates a T-shaped feature, as shown in Figs.~\ref{fig:phenomenon}(a)-(d). 
Furthermore, the eigenstates in the `$-$' horizontal part of the T-shaped spectrum display a marked distinction from those in the `$|$' vertical part. Specifically, the eigenstates in the `$-$' part localize at the left boundary, manifesting the NHSE. This is consistent with the fact that the `$-$' part of the periodic-boundary-condition (PBC) spectrum encircles the corresponding open-boundary-condition (OBC) spectrum \cite{PhysRevLett.124.086801,PhysRevLett.125.126402}, as shown in the inset of Figs.~\ref{fig:phenomenon}(a)-(d). Conversely, the eigenstates in the `$|$' part are localized around each site of chain B within the bulk region [see Fig.~\ref{fig:phenomenon}(e)]. Notably, the eigenstates in the`$-$' part predominantly reside in chain A [see Fig.~\ref{fig:phenomenon}(f)], while those in the `$|$' part in chain B [see Fig.~\ref{fig:phenomenon}(g)]. This implies that chains A and B are relatively independent.

\begin{figure}[t]
\includegraphics[width=8.6cm]{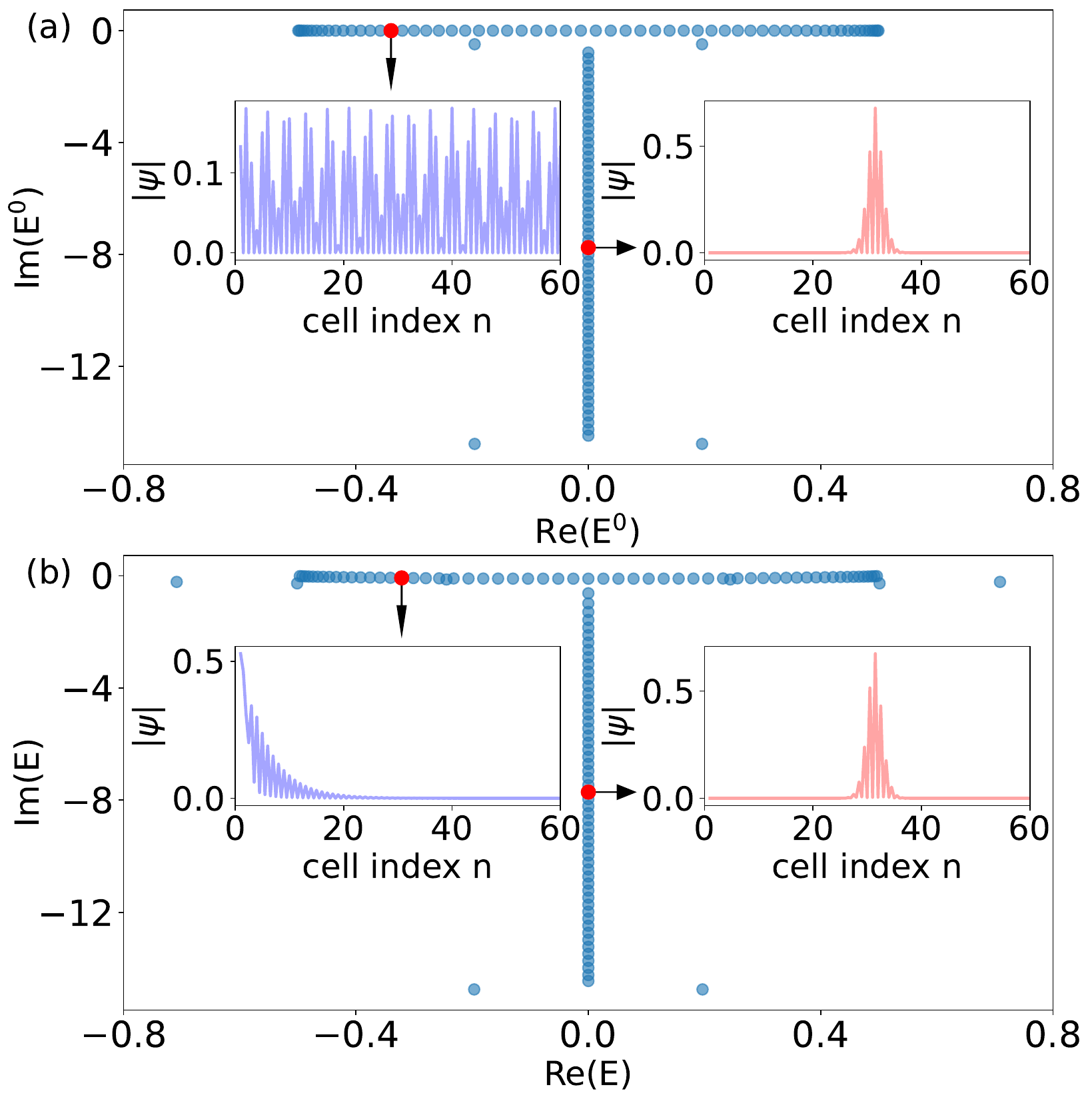}
\caption{\label{fig:t-shape} Energy spectra for turning `off' (a) and `on' (b) the coupling between chains A and B. Insets are the profiles of typical eigenstates from the `$-$' and `$|$' parts of the spectra, corresponding to the red points. The parameters are $t_1=0.4$, $t_2 = 0.5$, $\gamma_n = 0.25 n$, and $L = 60$.}
\end{figure}

To gain insights, we turn off the coupling between chains A and B, making them independent.
Hereafter, we use $E^0$ and $E$ to %label
denote the energies when chains A and B are decoupled and coupled, respectively. The energy of chain A is real and given by $E^{0, -} = t_2 \sin (k)$, whose eigenstates are extended standing waves with Bloch wave vector $k$.
In chain B, each site is linked to an eigenstate localized around it, with the energy closely approximated by its imaginary potential. This is similar to the Wannier-Stark localization \cite{ws_loc_1,ws_loc_2}. The superposition of the spectra of chains A and B results in a T-shaped spectrum, and the `$-$' and `$|$' parts are formed, as depicted in Fig.~\ref{fig:t-shape}(a), by the energies of chains A and B, respectively. Upon introducing the coupling between chains A and B, the spectrum undergoes certain distortions, but retains, almost unchanged, the T shape[see Fig.~\ref{fig:t-shape}(b)]. In such a coupled case, the eigenstates corresponding to the `$|$' part are still localized around their respective sites in chain B. However, the NHSE appears in the eigenstates %in 
of the `$-$' part. 

{\it ISSE.---}
For simplicity, we apply a local unitary transformation to the basis $\ket{n, A (B)}$ 
and then obtain a new basis $\ket{n,A^{\prime}(B^{\prime})}$.
Specifically, we define $\ket{n, A'} = \left( \ket{n, A} - i\ket{n, B} \right)/\sqrt{2}$ and $\ket{n, B'} = \left( \ket{n, A} + i\ket{n, B} \right)/\sqrt{2}$. Consequently, the eigenequations Eq.~(\ref{eqn:hamiltonian}) become more concise, yielding (more details in \cite{[{See Supplemental Material at }][{ for details.}]supp})
\begin{equation}\label{eqn:hamiltonian2}
\begin{aligned}
    E \psi_{n}^{A'} =& -i\frac{\gamma_n}{2}\psi_{n}^{A'} + t_2 \psi_{n-1}^{B'} + (t_1 + \frac{\gamma_n}{2})\psi_{n}^{B'}, 
    \\
    E \psi_{n}^{B'} =& -i\frac{\gamma_n}{2}\psi_{n}^{B'} + (t_1 - \frac{\gamma_n}{2})\psi_{n}^{A'} + t_2 \psi_{n+1}^{A'}, 
\end{aligned}
\end{equation}
where $\psi_n^{A'(B')} = \braket{n, A'(B')|\psi}$ is the wave function projected onto $\ket{n, A'(B')}$.

\begin{figure}[t]
    \includegraphics[width=8.6cm]{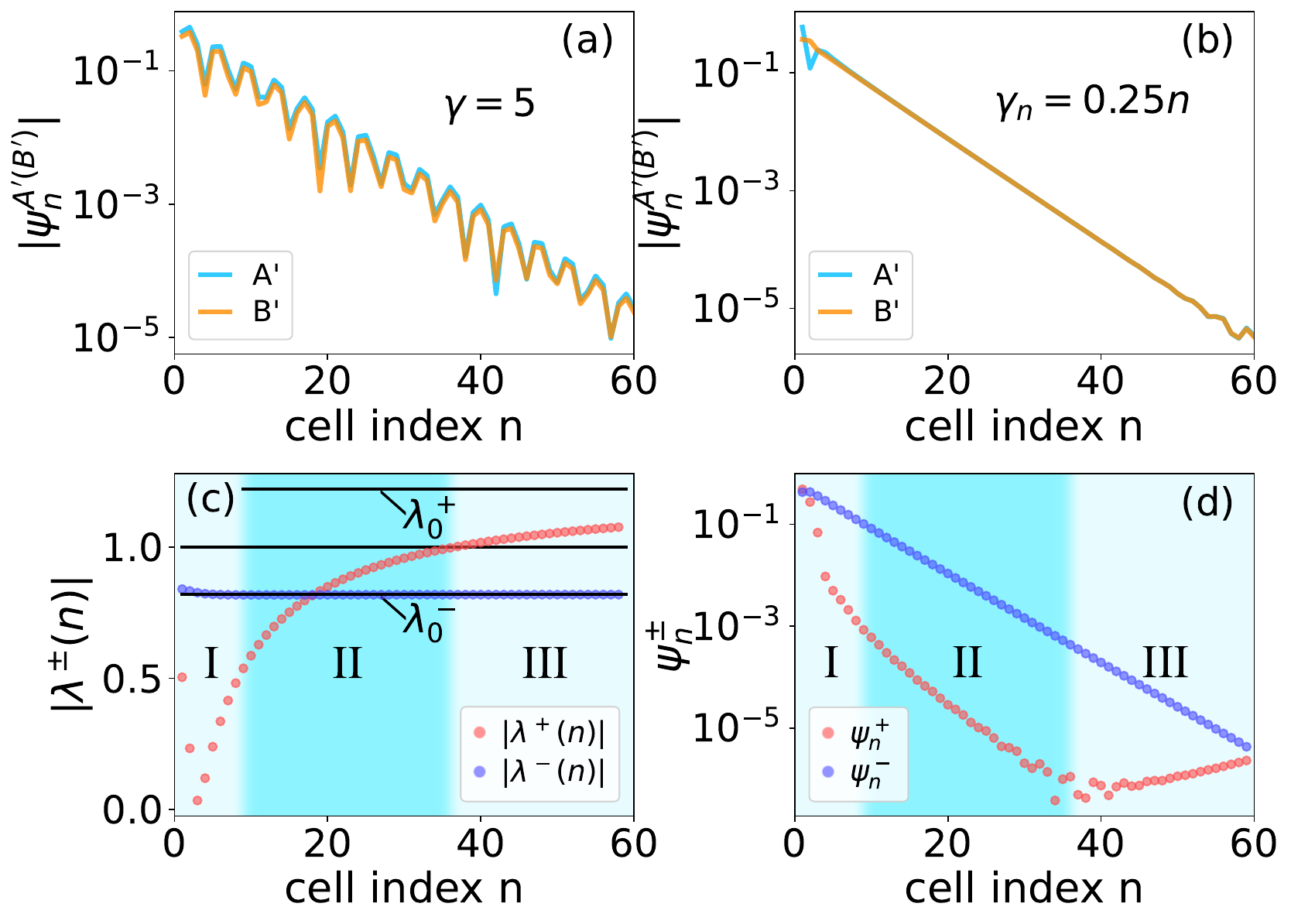}
    \caption{\label{fig:nhse} (a), (b) Profiles of the eigenstates for uniform and linear loss rates. For a uniform loss rate, two exponentially-decaying waves interfere in the bulk. Conversely, the linear loss case only exhibits a dominant exponentially-decaying and no interference. (c) Modulus of the eigenvalues of the transfer-matrix, $|\lambda^{\pm}(n)|$, versus $n$. The horizontal black lines from top to bottom are $\lambda^{+}_0$, $1$, $\lambda^{-}_0$, respectively. (d) Decomposition components of the wave function, $\psi^{\pm}_n$, versus $n$. The parameters are $t_1=0.4$, $t_2 = 0.5$, $L = 60$ in all plots, while $\gamma = 5$ in (a) and $\gamma_n = 0.25 n$ in (b)-(d).}
\end{figure}

The ISSE exhibits two features. 
First, although the loss rate $\gamma_n$ increases spatially in the lattice, the skin modes have an almost uniform decay rate within the bulk region. This is illustrated by the linearity of the wave function $\psi_{n}^{A^{\prime}(B^{\prime})}$ on a logarithmic scale [see Fig.~\ref{fig:nhse}(b)]. Second, for one-dimensional spatially-periodic tight-binding non-Hermitian models, 
the eigenstates can be written as a superposition of two exponentially-decaying waves. Under the OBCs, constructing continuum bands requires the amplitudes of these two waves to be equal \cite{prl_wang,PhysRevLett.123.066404,PhysRevLett.125.126402, PhysRevX.14.021011}, inducing interference in the bulk [see Fig.~\ref{fig:nhse}(a)].
However, in our model, numerical results show the absence of interference. This indicates that each skin mode can be approximately expressed as a single exponentially-decaying wave within the bulk region [see Fig.~\ref{fig:nhse}(b)], i.e.,
\begin{equation}\label{eqn:ansatz}
    \psi_n^{S'} \sim \beta^n,
\end{equation} 
where $S'\in \{A', B'\}$, and $\beta<1$ is a uniform decay rate. This result differs greatly from the spatially-periodic cases.

Let us first adopt a rough approach to obtain $\beta$ in Eq.~(\ref{eqn:ansatz}). 
We define the ratio $\beta_n = \psi_n^{S'}/\psi_{n-1}^{S'}$, and substitute $\beta_n$ into the bulk eigenequation in Eq.~(\ref{eqn:hamiltonian2}), yielding
\begin{equation}\label{eqn:8}
\begin{aligned}
0=& \gamma_n \left[ iE-\frac{t_2}{2}\left( \beta_{n+1}-\beta_{n}^{-1}\right)  \right]\\
&+ E^2 -\left(t_1 + t_2 \beta_{n}^{-1} \right)\left(t_1+t_2 \beta_{n+1} \right).
\end{aligned}
\end{equation}
To obtain solutions $\beta_n \approx \beta$ consistent with our numerical findings in Eq.~(\ref{eqn:ansatz}) and also to ensure this equation holds for arbitrary $n$,
the $\gamma_n$ term must be approximately zero, which gives two solutions of $\beta$:
\begin{equation}\label{eqn:beta}
    \beta_{\pm} \approx \frac{iE}{t_2} \left[ 1 \pm \sqrt{1-\left(t_2/E\right)^2} \right].
\end{equation}
Our simulation indicates that the analytical solution $\beta_-$ agrees with the numerical results of $\beta$.

{\it Transfer-matrix method.---} The above approach fails to explain why $\beta_-$, instead of $\beta_+$, fits $\beta$ and how the ISSE arises. Below, we analyze the model more accurately using the transfer-matrix method \cite{PhysRevLett.57.2057, PhysRevB.42.10329,transfer_matrix_2}. Consider the projected bulk eigenequations in a subspace spanned by the bases $\ket{n-1, B'} $ and $\ket{n, A'}$, which can be expressed, in terms of the transfer-matrix, as (more details in \cite{[{See Supplemental Material at }][{ for details.}]supp}):
\begin{equation} \label{eqn:tm}
    \ket{\psi(n)}
    = T(n) \ket{\psi(n-1)},
\end{equation}
where $\ket{\psi(n)} = (\psi_n^{A'},\psi_n^{B'})^\mathcal{T}$ is the wavefunction on the $n$-th unit cell, and $T(n)$ is an $2\times2$ transfer-matrix between the $(n-1)$- and $n$-th unit cells. Here, $\mathcal{T}$ refers to a transpose operation, and the $n$-th cell consists of the $n$-th sites of chains A and B.
If $\gamma_n$ is uniform, then $T(n)$  simplifies to a constant matrix $T_{0}$. As a result, $\ket{\psi(n)} = (T_{0})^{n-1} \ket{\psi(1)}$ reduces to the ansatz of non-Bloch band theory. In our model, $T(n)$ is a function of $n$, and its eigenequation can be written as:
\begin{equation}\label{eqn:eigeneqn}
    \lambda^2(n) + b(n)\lambda(n) + c(n) = 0.
\end{equation}
Here,
\begin{equation}\label{eqn:bc}
\begin{aligned}
    b(n)& = \frac{t_1^2+t_2^2-E^2+ \gamma_{n}^{-} t_1
           -i\gamma_n^{+} E}{ t_2\left(t_1+\gamma_n\right/2)},\\
    c(n)& = \frac{t_1-\gamma_{n-1}/2}{t_1+\gamma_n/2},
\end{aligned}
\end{equation} 
with $\gamma_{n}^{\pm}=(\gamma_{n}\pm\gamma_{n-1})/2$. It is clear that the two eigenvalues of the transfer-matrix $T(n)$ are $\lambda^{\pm}(n) = \left[ -b(n) \pm \sqrt{b^2(n)-4c(n)} \right]/2$, corresponding to the left and right eigenvectors, $\bra{\lambda^{\pm}_L(n)}$ and $\bra{\lambda^{\pm}_R(n)}$, which are orthonormal and complete.

Since $b(n)$ and $c(n)$  converge to $-2iE/t_2$ and $-1$, respectively, as $n$ approaches infinity, we therefore have
\begin{equation}\label{eqn:lambda0}
    \lim_{n\to\infty} \lambda^{\pm}(n) \to \lambda^{\pm}_0 = \frac{iE}{t_2} \left[ 1 \pm \sqrt{1-\left(t_2/E\right)^2} \right],
\end{equation}
which is the same as $\beta_{\pm}$ in Eq.~(\ref{eqn:beta}) obtained from the rough approach mentioned above. Recall that the `$-$' part of the T-shaped spectrum is $E^{0, -} = t_2 \sin (k)$ in the decoupled case of chains A and B. Thus, we define $\kappa = \arcsin(E/t_2)$, which is a complex number; then Eq.~(\ref{eqn:lambda0}) becomes $\lambda_0^{\pm} = \pm e^{\pm i \kappa}$ \footnote{Here we choose the principal branch of square root function in Eq.~(\ref{eqn:lambda0}) to ensure it is single-valued, by defining the angle of the argument of square root to be in $[0,\  2\pi )$}, which corresponds to the Bloch phase factor $\pm e^{\pm ik}$ in the decoupled case. It is easily seen that $|\lambda^{+}_0| > 1$ and $|\lambda^{-}_0| < 1$ \cite{[{See Supplemental Material at }][{ for details.}]supp}, consistent with the numerical results [see Fig.~\ref{fig:nhse}(c)].

Next, we examine the convergence of $\lambda^{\pm}(n)$ towards $\lambda^{\pm}_0$. As illustrated in Fig.~\ref{fig:nhse}(c), $\lambda^{-}(n)$ converges notably faster than $\lambda^{+}(n)$. Here, we provide a concise idea for the proof of this result. For convenience, we assume that $\gamma_n$ takes the form $\gamma_n = n \gamma_0 $. The detailed proofs for various types of $\gamma_n$ functions are given in \cite{[{See Supplemental Material at }][{ for details.}]supp}. To analyze the speed of convergence, we perform a Laurent expansion for $\lambda^{\pm}(n)$, given by
\begin{equation}\label{eqn:laurent}
    \lambda^{\pm}(n) = \lambda^{\pm}_0 + \lambda^{\pm}_1/\gamma_n + \cdots,
\end{equation}
and combining Eqs.~(\ref{eqn:eigeneqn}), (\ref{eqn:bc}) and (\ref{eqn:laurent}), we obtain
\begin{equation}\label{eqn:lambda_1_2}
    \frac{\lambda^{\pm}_1}{\lambda^{\pm}_0} = \mp \frac{(t_1 \pm t_2 \cos (\kappa) ) (t_1 \pm t_2 \cos (\kappa) + \gamma_0/2)}{t_2 \cos (\kappa)}.
\end{equation}
Equation (\ref{eqn:lambda_1_2}) characterizes the convergence speed of $ \lambda^{\pm}(n)$; that is, a smaller value of $\lambda^{\pm}_1/\lambda^{\pm}_0$ implies a smaller $1/\gamma_n$ term, and thus a faster convergence. In the thermodynamic limit, the `$-$' part of the spectrum in the coupled case of chains A and B closely approximates that in the decoupled case \cite{[{See Supplemental Material at }][{ for the details.}]supp}. Consequently,  $\kappa$ can be closely approximated by $k$ in the decoupled case, which ranges from $-\pi/2$ to $\pi/2$. Then, it follows from Eq.~(\ref{eqn:lambda_1_2}) that $|\lambda^{-}_1/\lambda^{-}_0|<|\lambda^{+}_1/\lambda^{+}_0|$. In particular, $|\lambda^{-}_1/\lambda^{-}_0|$ approaches zero when $\cos(\kappa)$ is close to $t_1/t_2$ or $(t_1 + \gamma_0/2)/t_2$, leading to a much faster convergence speed of $\lambda^{-}(n)$.

We can now analyze the formation of the ISSE states in our model. We decompose $\ket{\psi(n)}$ into the eigenvectors of $T(n)$:
\begin{equation}\label{eqn:psi^pm_n}
    \ket{\psi(n)} = \psi_n^{+} \ket{\lambda^{+}_R(n)} + \psi_n^{-} \ket{\lambda^{-}_R(n)},
\end{equation}
where $\psi_n^{\pm} = \braket{\lambda^{\pm}_L(n) | \psi(n)}$.
If $\gamma_n$ is uniform, then $\lambda^{\pm}(n) = \lambda^{\pm}$ is constant. Together with Eq.~(\ref{eqn:tm}), we obtain $\psi_n^{\pm} = \lambda^{\pm}\psi_{n-1}^{\pm}$, which aligns with the intuition that $\psi_n^{\pm}$ are scaled by the corresponding eigenvalues $\lambda^{\pm}$. When returning to the nonuniform $\gamma_n$ and assuming that the transfer-matrix $T(n)$ varies slowly, we can take the approximation \cite{[{See Supplemental Material at }][{ for the details.}]supp}
\begin{equation}\label{eqn:coef}
    \psi_n^{\pm} \approx \lambda^{\pm}(n) \psi_{n-1}^{\pm}.
\end{equation}

Figure \ref{fig:nhse}(d) shows how $\psi_n^{\pm}$ evolves with $n$. The components of $\ket{\psi(n)}$, namely
$\psi_n^{+}$ and $\psi_n^{-}$, need to be of the same order at the two edges, so that they can cancel each other to satisfy the OBCs $\ket{\psi(0)} = \ket{\psi(L+1)} = 0$.  We divide the wave function $\psi_{n}^{\pm}$ into three regions, labeled by I, II, and III, as in Figs.~\ref{fig:nhse}(c) and \ref{fig:nhse}(d). These three regions can be roughly viewed as the left-boundary, bulk, and right-boundary regions, respectively. In region I, near the left edge, $\psi_n^{+}$ and $\psi_n^{-}$ have the same order. The eigenvalue $\lambda^{-}(n)$ converges to $\lambda^{-}_0$ quickly, so according to Eq.~(\ref{eqn:coef}), $\psi_n^{-} \sim \left(\lambda^{-}_0\right)^n$ behaves like an exponentially-decaying wave. Moreover, as $\gamma_n$ increases from 0 to $2t_1$, $\lambda^{+}(n) = c(n)/\lambda^{-}(n) \approx c(n)/\lambda^{-}_0 $ approaches $0$. Hence, $|\psi_n^{+}|$ decays much faster than $|\psi_n^{-}|$, ultimately reaching an extremely low value. In region II, $|\psi_n^{+}|$ is several orders of magnitude lower than $|\psi_n^{-}|$, so $\psi_n^{-}$ dominates the behavior of the state $\ket{\psi(n)}$ and has a rigorously exponential decay.
As $n$ grows, $|\lambda^{+}(n)|$ monotonically increases to $|\lambda^{+}_0|$. When $|\lambda^{+}(n)|>1$, it signifies a transition to region III, near the right boundary. In this region, $|\psi_n^{+}|$ starts to increase, and ultimately approaches the order of $|\psi_n^{-}|$ at the right edge, to satisfy the OBCs.
Examining these regions reveals that $|\psi_n^{-}|$ is much larger than $|\psi_n^{+}|$ in the bulk. Therefore, the $\psi_n^{-} \ket{\lambda^{-}_R(n)}$ term of Eq.~(\ref{eqn:psi^pm_n}) dominates the behavior of $\ket{\psi(n)}$, which scales as $\left(\lambda^{-}_0\right)^n$. 
In contrast, the $\psi_n^{+} \ket{\lambda^{+}_R(n)}$ term becomes dominant only near the boundaries, thus causing fluctuations in the wave function at these areas [see Fig.~\ref{fig:nhse}(b)]. 

Our analysis shows that the formation of the ISSE in our model demands a notably fast convergence speed of $\lambda^{-}(n)$, leading to a single stable exponentially-decaying wave that predominantly governs the behavior of the skin modes in the bulk. We also discuss the impact of parameters in \cite{[{See Supplemental Material at }][{ for the details.}]supp}. An extremely small or large $t_1$, compared to $t_2$, can result in a slow convergence speed of $\lambda^{-}(n)$, which weakens the features of the ISSE. This also supports the relationship between the ISSE and the convergence speed of $\lambda^{-}(n)$. Furthermore, we demonstrate that it is possible to show the ISSE in a short lattice, which provides the feasibility of future experimental investigations with finite-size systems \cite{[{See Supplemental Material at }][{ for the details.}]supp}.

{\it Conclusions.---} We unveil a novel class of NHSE, the ISSE, arising in a nonuniform lossy lattice. Unlike the conventional NHSE described by non-Bloch band theory, the ISSE exhibits a single dominant exponentially-decaying wave within the bulk. Such a peculiar behavior is closely related to the convergence speed of the transfer-matrix eigenvalues. The eigendecomposition of the transfer-matrix reveals that the wave function comprises two parts: a dominant, exponentially-decaying component that governs bulk behavior, and a negligible component that is impactful only at the boundaries. Our work provides a new perspective for accurately quantifying the NHSE in translational symmetry-breaking systems and can be extended to other models, e.g., Floquet systems with nonuniform dissipation \cite{annurev-conmatphys-040721-015537,PhysRevB.108.L220301}, interacting non-Hermitian system and\cite{PhysRevLett.121.203001, PhysRevLett.129.180401, Li2024Dissipation} and Liouvillian skin effect\cite{PhysRevLett.127.070402, PhysRevResearch.4.023160, Kuo2024Non}.

\begin{acknowledgments}
The authors would like to thank professor Jieqiao Liao, doctor Pengyu Wen and Chenyang Wang for helpful discussion. 
This work is supported by the National Natural Science Foundation of China under Grants No. 11974205, and No. 61727801, and the Key Research and Development Program of Guangdong province (2018B030325002). 
W.Q.~acknowledges support of the National Natural Science Foundation of China (NSFC) via Grant No.~0401260012. 
F.N. is supported in part by:
the Japan Science and Technology Agency (JST)
[via the CREST Quantum Frontiers program Grant No. JPMJCR24I2,
the Quantum Leap Flagship Program (Q-LEAP), and the Moonshot R\&D Grant Number JPMJMS2061],
and the Office of Naval Research (ONR) Global (via Grant No. N62909-23-1-2074).
\end{acknowledgments}

\bibliography{Imaginary_Stark_Skin_Effect}% Produces the bibliography via BibTeX.

\end{document}

\preprint{APS/123-QED}

\title{Supplemental Material for ``Imaginary-Stark Skin Effect"}

\author{Heng Lin}
    \affiliation{State Key Laboratory of Low-Dimensional Quantum Physics and Department of Physics, Tsinghua University, Beijing 100084, China}
\author{Jinghui Pi}
    \email{pijh14@gmail.com}
    \affiliation{State Key Laboratory of Low-Dimensional Quantum Physics and Department of Physics, Tsinghua University, Beijing 100084, China}
\author{Yunyao Qi}
    \affiliation{State Key Laboratory of Low-Dimensional Quantum Physics and Department of Physics, Tsinghua University, Beijing 100084, China}
\author{Wei Qin}   
    \affiliation{Center for Joint Quantum Studies and Department of Physics, Tianjin University, Tianjin 300350, China}
\author{Franco Nori}
     \affiliation{Center for Quantum Computing, RIKEN, Wako-shi, Saitama 351-0198, Japan}
    \affiliation{Department of Physics, The University of Michigan,
	Ann Arbor, Michigan 48109-1040, USA}
\author{Gui-Lu Long}%
    \email{gllong@tsinghua.edu.cn}
    \affiliation{State Key Laboratory of Low-Dimensional Quantum Physics and Department of Physics, Tsinghua University, Beijing 100084, China}
    \affiliation{Frontier Science Center for Quantum Information, Beijing 100084, China}
    \affiliation{Beijing Academy of Quantum Information Sciences, Beijing 100193, China}
             
\maketitle

\hypersetup{linkcolor=blue}
\tableofcontents
\hypersetup{linkcolor=red}

% The Supplemental Material contains technical details regarding: (i) Derivation of transfer matrix; (ii) The modulus of $\lambda^{\pm}_0$; (iii) Energy change between A-B decoupled and coupled systems; (iv) Convergence property of transfer matrix eigenvalues $\lambda^{\pm}(n)$; (v) The recurrence relation of $\psi_n^{\pm}$ and its approximation; (vi) Analysis of Parameters.

\section{Rotation transformation on the Hamiltonian}
The Hamiltonian of our model is given by
\begin{equation}
\begin{aligned}
    \hat{H} =& i\frac{t_2}{2} \sum_{n=1}^{L-1} \Big[\ket{n+1,A}\bra{n,A} - \ket{n+1,B}\bra{n,B} \Big] + \mathrm{h.c.} \\
      & + \frac{t_2}{2} \sum_{n=1}^{L-1} \Big[ \ket{n+1,B}\bra{n,A} + \ket{n+1,A}\bra{n,B} \Big] + \mathrm{h.c.} \\
      & + t_1 \sum_{n=1}^{L} \ket{n,A}\bra{n,B} + \mathrm{h.c.}  - \sum_{n=1}^{L} i\gamma_n \ket{n,B}\bra{n,B}.
\end{aligned}
\end{equation}
We regard each unit cell as a pseudo spin. Specifically, we define $\ket{n,A}$ as $\ket{\sigma_z^n = +\frac{1}{2}}$ and $\ket{n,B}$ as $\ket{\sigma_z^n = -\frac{1}{2}}$. Then, we apply a $\pi/2$ rotation transformation along the $x$-axis to each spin, i.e.,
\begin{equation}
    \hat{H'} = \mathcal{R}^{-1} \hat{H} \mathcal{R}, 
\end{equation}
where $\mathcal{R}$ is the spin rotation operator given by
\begin{equation}
    \mathcal{R} =  \bigoplus_n \exp\left(-i \frac{\pi}{4} \sigma_x^n\right).
\end{equation}
After the rotation, the Hamiltonian becomes
\begin{equation}
\begin{aligned}
    \hat{H'} =& t_2 \sum_{n=1}^{L-1} \ket{n+1, A'}\bra{n, B'} + \mathrm{h.c.}\\
       & + \sum_{n=1}^{L}\Big[ (t_1+\frac{\gamma_n}{2})\ket{n,A'}\bra{n,B'} + (t_1-\frac{\gamma_n}{2})\ket{n,B'}\bra{n,A'} \Big]\\
       &  -i\frac{\gamma_n}{2}\sum_{n=1}^{L} \Big[ \ket{n,A'}\bra{n,A'} + \ket{n,B'}\bra{n,B'} \Big],
\end{aligned}
\end{equation}
where $\ket{n,A'}$ and $\ket{n,B'}$ are the new Z bases of the pseudo spin after the rotation. The transformation of the wave function is given by
\begin{equation}
    \begin{pmatrix}
        \psi_n^{A'} \\
        \psi_n^{B'}
    \end{pmatrix}
    = \mathcal{R}^{-1}
    \begin{pmatrix}
        \psi_n^A \\
        \psi_n^B
    \end{pmatrix},
\end{equation}
that is, 
\begin{equation}
\begin{aligned}
    &\psi_n^{A'} = \frac{\sqrt{2}}{2} \psi_n^{A} + i \frac{\sqrt{2}}{2} \psi_n^B, \\
    &\psi_n^{B'} = i \frac{\sqrt{2}}{2} \psi_n^{A} + \frac{\sqrt{2}}{2} \psi_n^B.
\end{aligned}
\end{equation}

\section{Derivation of the transfer-matrix}
In this section, we provide a detailed derivation of the transfer-matrix.
Consider the bulk eigen-equations projected onto the bases $\ket{n-1, B'} $ and $\ket{n, A'}$: 
$\bra{n-1, B'}H'\ket{\psi} = \bra{n-1, B'}E\ket{\psi}$ and $\bra{n, A'}H'\ket{\psi} = \bra{n, A'}E\ket{\psi}$,
which can be expressed as
\begin{equation}
    \begin{pmatrix}
        t_1 - \frac{\gamma_{n-1}}{2} & -i\frac{\gamma_{n-1}}{2}-E & t_2 & 0 \\
        0 & t_2 & -i\frac{\gamma_n}{2}-E & t_1 + \frac{\gamma_n}{2}
    \end{pmatrix}
    \begin{pmatrix}
        \psi_{n-1}^{A'} \\ \psi_{n-1}^{B'} \\ \psi_n^{A'} \\ \psi_n^{B'}
    \end{pmatrix}
    =0.
\end{equation}
This can be rewritten as
\begin{equation}
    \begin{pmatrix}
        t_2 & 0 \\
        -i\frac{\gamma_n}{2}-E & t_1 + \frac{\gamma_n}{2}
    \end{pmatrix}
    \begin{pmatrix}
        \psi_n^{A'} \\ \psi_n^{B'}
    \end{pmatrix}
    =-
    \begin{pmatrix}
        t_1 - \frac{\gamma_{n-1}}{2} & -i\frac{\gamma_{n-1}}{2}-E \\
        0 & t_2
    \end{pmatrix}
    \begin{pmatrix}
        \psi_{n-1}^{A'} \\ \psi_{n-1}^{B'}
    \end{pmatrix},
\end{equation}
which directly leads to
\begin{equation} \label{eqn:tm}
    \ket{\psi(n)}
    = T(n) \ket{\psi(n-1)},
\end{equation}
where $\ket{\psi(n)} = (\psi_n^{A'},\ \psi_n^{B'})^\mathcal{T}$, and $T(n)$ is the transfer-matrix between cell $(n-1)$ and cell $n$, given by
\begin{equation} \label{eqn:tn}
\begin{aligned}
    T(n) 
    =& -
    \begin{pmatrix}
        t_2 & 0 \\
        -i\frac{\gamma_n}{2}-E & t_1 + \frac{\gamma_n}{2}     
    \end{pmatrix}^{-1}
    \begin{pmatrix}
        t_1 - \frac{\gamma_{n-1}}{2} & -i\frac{\gamma_{n-1}}{2}-E \\
        0 & t_2
    \end{pmatrix}\\
    =& \begin{pmatrix}
        -\frac{t_1-\gamma_{n-1}/2 }{t_2}  &  
        \frac{E+i\gamma_{n-1}/2}{t_2} \\
        -\frac{(E+i \gamma_{n}/2)(t_1-\gamma_{n-1}/2)}{t_2(t_1+ \gamma_{n}/2)} & 
        \frac{(E+i \gamma_{n-1}/2)(E+i \gamma_{n}/2)-t_2^2}{t_2(t_1+\gamma_{n}/2)}
    \end{pmatrix}.
    \end{aligned}
\end{equation}

\section{The modulus of the parameters $\lambda^{\pm}_0$}

Let us begin with the following conclusion from the main text:
\begin{equation}
    \lambda_0^{\pm} = \pm \exp(\pm i \kappa),
\end{equation}
where $\kappa = \arcsin(E/t_2)$ is a complex number. Given that our system is dissipative, the imaginary part of the eigenenergy is negative, i.e., $\operatorname{Im}(E) < 0$. Therefore, the imaginary part of $\kappa$, namely $\operatorname{Im}(\kappa)$, is also negative, indicating that
\begin{equation}
    |\lambda^{+}_0| = \exp\left[-\operatorname{Im}(\kappa)\right] > 1, \quad \quad \quad 
    |\lambda^{-}_0| = \exp\left[\operatorname{Im}(\kappa)\right] < 1.
\end{equation}

\section{Energy difference between the decoupled and coupled  configurations of chains A and B}\label{section1}

In this section, we show that the energy difference in the `$-$' horizontal part of the T-shaped spectrum, between the decoupled and coupled configurations of chains A and B, vanishes in the thermodynamic limit. For clarity, we introduce the following notations for these two system configurations. The Hamiltonian in the decoupled case is denoted as $H^{0}$, while the Hamiltonian in the coupled case is represented as $H$. The difference between these Hamiltonians, denoted by $H'$, is defined as
$H' \defeq H - H^{0}$. The right eigenstates of  $H^{0}$ are denoted by $ \ket{R^{0}_i} $ with corresponding eigenenergies $ E^{0}_i$, where the subscript $i$ is the index of the eigenstates. Furthermore, we specify $E^{0, -}_i$ ($E^{0, |}_i$) and $\ket{R^{0, -}_i}$ ($\ket{R^{0, |}_i}$) as the eigenenergies and eigenstates belonging to the `$-$' (`$|$') subspace of the $H^{0}$ spectrum, respectively. We use similar notations $E_i$, $E^{-}_i$, and $E^{|}_i$ to denote the corresponding energy spectra in the coupled case.

In the decoupled case of chains A and B, for the `$-$' part of the spectrum, the eigenenergies are given by $E^{0, -}_i = t_2 \sin k$, where $k = \left( \frac{i}{L+1} - \frac{1}{2}\right)\pi$ and $i = 1, 2, \ldots, L$.  The corresponding eigenstates are the extended standing waves in the bulk, expressed as $\ket{R^{0, -}_i} = \sqrt{\frac{2}{L}}\sum_{n=1}^L {(-1)^{n/2} \sin\left[(\frac{\pi}{2} - k)n\right] \ket{n, A}}$. 
For the `$|$' vertical part of the spectrum, the eigenstates are localized around the sites of chain B, with energies that are approximately equal to $-i\gamma_n$.

\begin{figure}[b]
\includegraphics[width=0.7\textwidth]{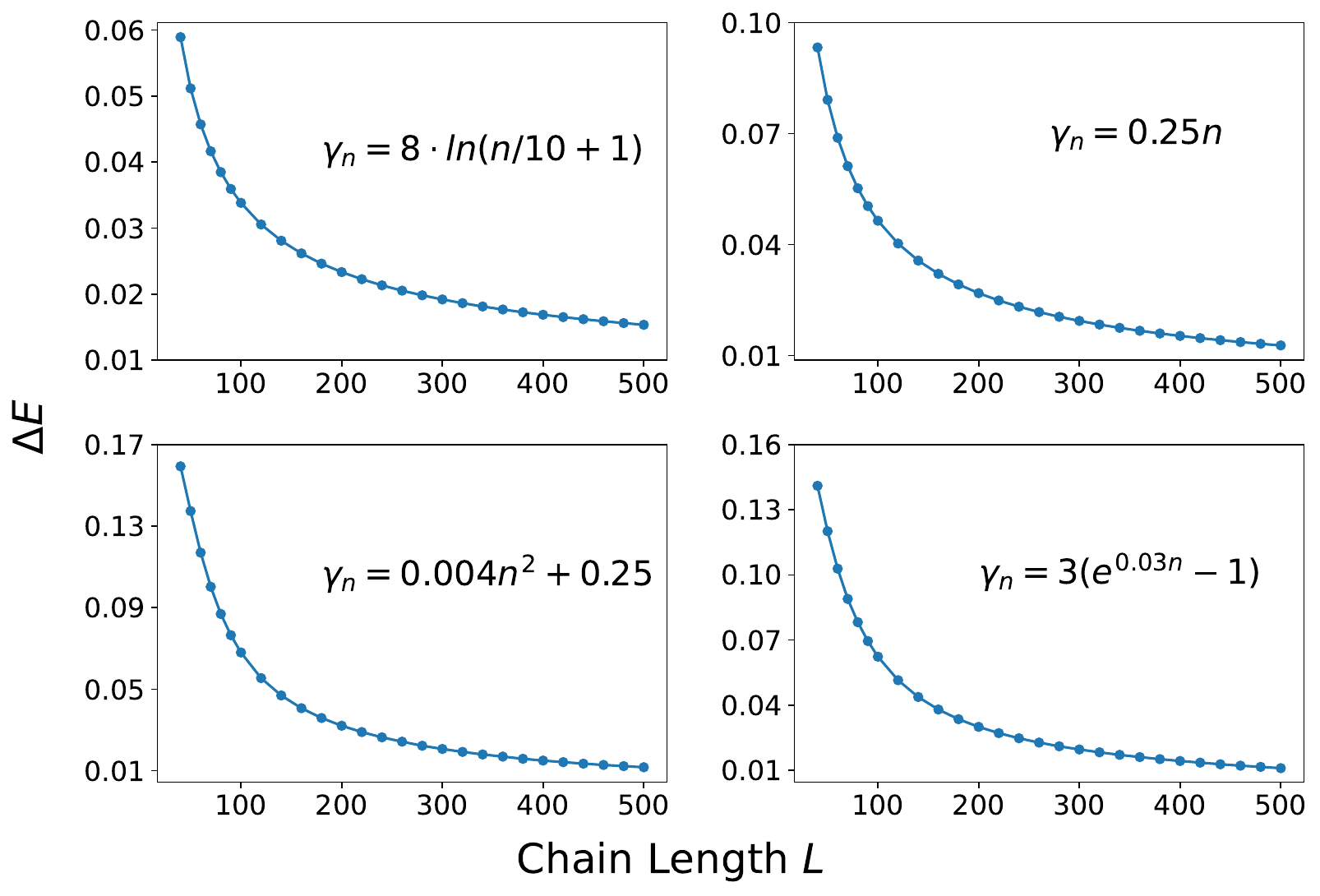}
\caption{\label{fig:DeltaE} Average energy difference in the `$-$' part of the spectrum between the decoupled and coupled cases, given by $\Delta E = \left(N^{0, -}\right)^{-1}\sum_{i} {|E^{0, -}_i - E^{-}_i|}$, where $N^{0, -} = \left| \left\{ E^{0, -}_i \right\}_i \right|$ denotes the number of eigenenergies in the `$-$' part of the spectrum. This difference is depicted
as a function of chain length $L$ for logarithmic, linear, quadratic, and exponential loss rate functions. The parameters are $t_1=0.4$ and $t_2 = 0.5$.}
\end{figure}

Now we focus on the energy difference in the `$-$' part, between the decoupled and coupled cases of chains A and B, based on the following assumption:
\begin{assumption}
    Define $H(t) \defeq H^{0} + tH'$ for $t\in[0,1]$, such that $H(0) = H^{0}$ and $H(1) = H$. Consider an eigenvalue in the `$-$' part of the spectrum of $H^{0}$, denoted as $E^{0, -}_i$, along with its corresponding eigenvalue $E^{-}_i$ in the `$-$' part of the spectrum of $H$. 
    As the parameter $t$ varies, the eigenvalue of $H(t)$ traces a trajectory, denoted by $E_i(t)$. 
    This trajectory connects $E^{0}_i$ to $E_i$, meaning that $E_i(0) = E^{0, -}_i$ and $E_i(1) = E^{-}_i$,  without any level crossings.
\end{assumption}
This assumption is akin to the adiabatic assumption in Hermitian systems.  Based on this assumption, we argue that the energy from the `$-$' part in the decoupled case of chains A and B converges to that in the coupled cases; namely $E^{0, -}_i \sim E^{-}_i$, in the thermodynamic limit. 

In Fig.~\ref{fig:DeltaE}, we show that the average energy difference $\Delta E = \overline{|E^{0, -}_i - E^{-}_i|}$ decreases as the chain length $L$ increases. This reduction in $\Delta E$ can be attributed to two primary factors. First, as $L$ increases, the average separation between energies from the `$-$' part and the `$|$' part also increases, scaling as $\gamma_n$. Second, consider the strength of the couplings, introduced by the perturbation $H'$, between the state $\ket{R^{0, -}_i}$ and the eigenstates from the `$|$' part. Here, the eigenstates from the `$|$' part are localized states around the sites of chain B, denoted by $(n, B)$, and $H'$ is a local coupling. Due to this localized nature, through $H'$, the eigenstates from the `$|$' part can only couple with the portion of the wave function that is near $(n,B)$. However, the state $\ket{R^{0, -}_i}$ is an extended standing wave, with an amplitude that diminishes with $L$, scaling as $\sqrt{1/L}$. Therefore, the coupling between the state $\ket{R^{0, -}_i}$ and the eigenstates from the `$|$' part weakens as $L$ increases. These two factors collectively contribute to the attenuation of the virtual process between the eigenstates from the `$-$' part and the `$|$' part, resulting in a diminishing energy perturbation.

\begin{figure}[b]
\includegraphics[width=0.5\textwidth]{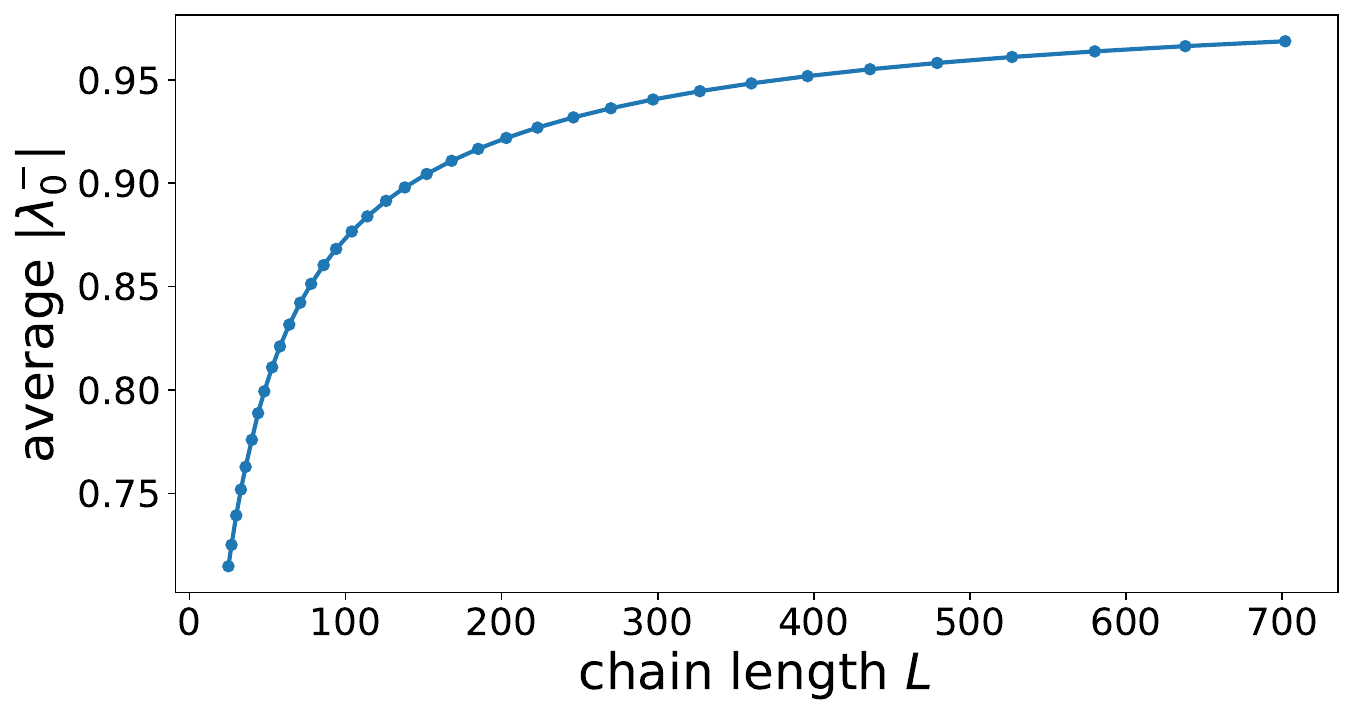}
\caption{\label{fig:decay_rate_vs_L} Average of $|\lambda_0^{-}|$ over eigenstates in the `$-$' part of the spectrum as a function of chain length $L$. The average $|\lambda_0^{-}|$ increases monotonically and tends to approach 1 as $L$ increases.The parameters are $t_1 = 0.4$, $t_2 = 0.5,$ and $\gamma_n = 0.25n$.}
\end{figure}

To elaborate further, we can use eigenvalue perturbation theory to analyze this model, by treating
$H'$ as the perturbation acting on $H^{0}$ \footnote{The convergence of eigenvalue perturbation series is a subtle problem in quantum mechanics, with many well-known cases exhibiting divergence. Nevertheless, our focus here is on a system with a finite chain length $L$ and a correspondingly finite dimension of its Hilbert space. This finite dimensionality ensures the convergence of perturbation theory in this specific scenario \cite{perturbation1982}. Subsequently, we delve into an analysis of how the perturbed eigenvalues evolve  as $L$ increases.}. Since $H'$ only couples adjacent A and B sites, the first-order perturbation of $E^{0, -}_i$ is zero. Instead, the dominant perturbation term arises from the second-order process between the eigenstate of $E^{0, -}_i$ and the eigenstates of the `$|$' part spectrum, represented by 
\begin{equation}\label{eqn:second-order}
    E^{2,-}_i = \sum_{j} {\frac{\braket{L^{0, -}_i | H' | R^{0, |}_j}\braket{L^{0, |}_j | H' | R^{0, -}_i}}{E^{0,-}_i - E^{0, |}_j}},
\end{equation}
where $\bra{L^{0, -}_i}$ and $\bra{L^{0, |}_j}$ are the left eigenstates corresponding to $E^{0, -}_i$  and $E^{0, |}_j$, respectively. In the term $\braket{L^{0, -}_i | H' | R^{0, |}_j}$ of Eq.~(\ref{eqn:second-order}), the state $\ket{R^{0, |}_j}$ is a localized wave function with a width of $ w(j) \sim w_0/\gamma_j' $, where $w_0$ is a constant and $\gamma_j' = \left. \frac{\mathrm{d} \gamma_n}{\mathrm{d}n}\right|_{n=j}$. This phenomenon is similar to the Wannier-Stark localization \cite{ws_loc_1,ws_loc_2}. Since $H'$ only couples adjacent A and B sites, $H'\ket{R^{0, |}_j}$ also yields a localized wave function with a width approximately equal to $w(j)$. Thus, its norm satisfies
\begin{equation}
    \left|H'\ket{R^{0, |}_j}\right| \leq (t_1+t_2)\left|\ket{R^{0, |}_j}\right| = t_1+t_2,
\end{equation}
where $(t_1+t_2)$ is the largest eigenvalue of $H'$. In contrast, the state $\bra{L^{0, -}_i} = \sqrt{\frac{2}{L}}\sum_{n=1}^L {(-1)^{n/2} \sin\left[(\frac{\pi}{2} - k)n\right] \bra{n, A}}$ represents a standing wave, which is extended in the bulk. As a result, the matrix element $\braket{L^{0,-}_i | H' | R^{0, |}_j}$ is bounded by
\begin{equation}
    \left|\braket{L^{0, -}_i | H' | R^{0, |}_j}\right| \leq (t_1+t_2)\sqrt{\frac{2}{L}}w(j)  = \mathcal{O} \left( \frac{1}{\gamma_j'\sqrt{L}} \right),
    \label{ineq3}
\end{equation}
where $\mathcal{O}$ denotes the big O notation, which describes the limiting behavior of a function.
For the denominator $\left(E^{0, -}_i - E^{0, |}_j\right)$ in Eq.~(\ref{eqn:second-order}), since $\left|E^{0, |}_j\right| \gg \left|E^{0, -}_i\right|$, we can derive the inequality:
\begin{equation}
    \left| E^{0, -}_i - E^{0, |}_j \right| \ge \left|E^{0, |}_j\right| - \left|E^{0, -}_i\right|  \ge \gamma_j - t_2 - t_2 = \gamma_j - 2t_2.
    \label{ineq4}
\end{equation}
By combining inequalities (\ref{ineq3}) and  (\ref{ineq4}), the upper bound of $E^{2,-}_i$ can be expressed as
\begin{equation}\label{eqn:uooer_bound}
    E^{2,-}_i \le \frac{C}{L} \sum_j \frac{1}{{\gamma_j'}^2 (\gamma_j - 2t_2)} = \mathcal{O} \left( \frac{1}{L} \sum_j \frac{1}{{\gamma_j'}^2 \gamma_j} \right),
\end{equation}
where $C$ is a relevant constant.
For a polynomial loss rate $\gamma_n = n^{\alpha}$, with $\alpha \ge 1$, or an exponential loss rate, it follows that
\begin{equation}
    \lim_{L\to\infty}{\frac{1}{L} \sum_j \frac{1}{{\gamma_j'}^2 \gamma_j}} = 0,
\end{equation}
which implies that the second-order perturbation of the energy $E^{2,-}_i$ converges to zero in the thermodynamic limit. Although this upper bound diverges for a logarithmic loss rate, numerical results suggest that, for a logarithmic loss rate function, the average energy difference in the `$-$' part between the  decoupled and coupled configurations of chains A and B also decreases as the chain length $L$ increases (see Fig.~\ref{fig:DeltaE}).

Upon $E^{-}_i \sim E^{0, -}_i$, one might raise the question of why a NHSE state and an extended state share similar energy values. In fact, as the parameter $L$ increases, the NHSE weakens, and the decay rate $\lambda^{-}_0$ of the wave function tends toward $1$, as illustrated in Fig.~\ref{fig:decay_rate_vs_L}. This makes the wave function more akin to an extended state. This phenomenon aligns with the physical intuition that a larger lattice implies a greater average loss rate in this model. Additionally, a lattice with a uniform loss rate will exhibit a weaker NHSE when the loss rate is larger.

\section{Convergence of the transfer-matrices eigenvalues $\lambda^{\pm}(n)$}\label{sec3}
In the main article, we state that $\lambda^{-}(n)$ converges much faster than $\lambda^{+}(n)$. In this section,  we delve into the convergence properties of $\lambda^{\pm}(n)$ in detail and establish the validity of this assertion across a broad range of types of the function $\gamma_n$.

We begin with the eigenequation of the transfer-matrices $T(n)$, given by
\begin{equation}\label{eqn:eigeneqn}
    \lambda^2 (n)+ b(n)\lambda(n) + c(n) = 0,
\end{equation}
where the expressions for $b(n)$ and $c(n)$ are
\begin{equation}\label{eqn:bc}
\begin{aligned}
   b(n)& = \frac{t_1^2+t_2^2-E^2+ \gamma_{n}^{-} t_1
           -i\gamma_n^{+} E}{ t_2\left(t_1+\gamma_n\right/2)},\\
    c(n)& = \frac{t_1-\gamma_{n-1}/2}{t_1+\gamma_n/2}.
\end{aligned}
\end{equation}
Here, $\gamma_{n}^{\pm}=(\gamma_{n}\pm\gamma_{n-1})/2$.
The solutions of Eq.~(\ref{eqn:eigeneqn}) can be expressed as
\begin{equation}\label{eqn:lambda}
    \lambda^{\pm}(n) = \frac{1}{2}\left( -b(n) \pm \sqrt{b^2(n)-4c(n)} \right).
\end{equation}
To ensure the single-valuedness of Eq.~(\ref{eqn:lambda}), we utilize the principal branch of the square root function, restricting the angle of the square root's argument to the interval $[ 0,\  2\pi )$. 

According to Eq.~(\ref{eqn:bc}), $b(n)$ and $c(n)$ converge to the constant values $b_0$ and $c_0$, respectively, as $n$ approaches infinity. For example, we have:
\begin{equation}
    \lim_{n\to\infty} b(n) \to b_0 =  -2iE/t_2,\quad \quad \quad \lim_{n\to\infty}  c(n) \to c_0 =  -1,
\end{equation}
for the polynomial loss rate $\gamma_n = n^{\alpha}$. Consequently,  $\lambda^{\pm}(n)$ converge to:
\begin{equation}\label{eqn:lambda_0}
    \lim_{n\to\infty} \lambda^{\pm}(n) \to \lambda^{\pm}_0 = \frac{1}{2}\left( -b_0 \pm \sqrt{b_0^2-4c_0} \right) = \frac{iE}{t_2} \left[ 1 \pm \sqrt{1-\left(t_2/E\right)^2} \right]. 
\end{equation}
To analyze the convergence properties of $\lambda^{\pm}(n)$, we expand Eq.~(\ref{eqn:bc}) in terms of  $1/\gamma_n$, yielding:
\begin{equation}\label{eqn:laurent}
\begin{aligned}
    b(n) &= b_0 + \frac{\gamma_n - \gamma_{n-1}}{\gamma_n} \frac{t_1 + iE}{t_2} + 
           \frac{1}{\gamma_n} \frac{2(t_1^2+t_2^2-E^2+2iEt_1)}{t_2} + \frac{b_2}{\gamma_n^2} + \cdots, \\
    c(n) &= c_0 + \frac{\gamma_n - \gamma_{n-1}}{\gamma_n} + \frac{4t_1}{\gamma_n} + \frac{c_2}{\gamma_n^2} + \cdots,
\end{aligned}
\end{equation}
where $b_2$ and $c_2$ are the coefficients of the $\left(1/\gamma_n\right)^2$ terms. 
To isolate the leading-order contribution,
we neglect higher-order terms $\left(1/\gamma_n\right)^m$ for $m >1$, retaining only the $\frac{1}{\gamma_n}$ and $\frac{\gamma_n - \gamma_{n-1}}{\gamma_n}$ terms. The specific form of $\gamma_n$ determines which one is the leading nontrivial term. We will discuss several different types of the function $\gamma_n$  in the following.

\begin{figure}[b]
\includegraphics[width=1\textwidth]{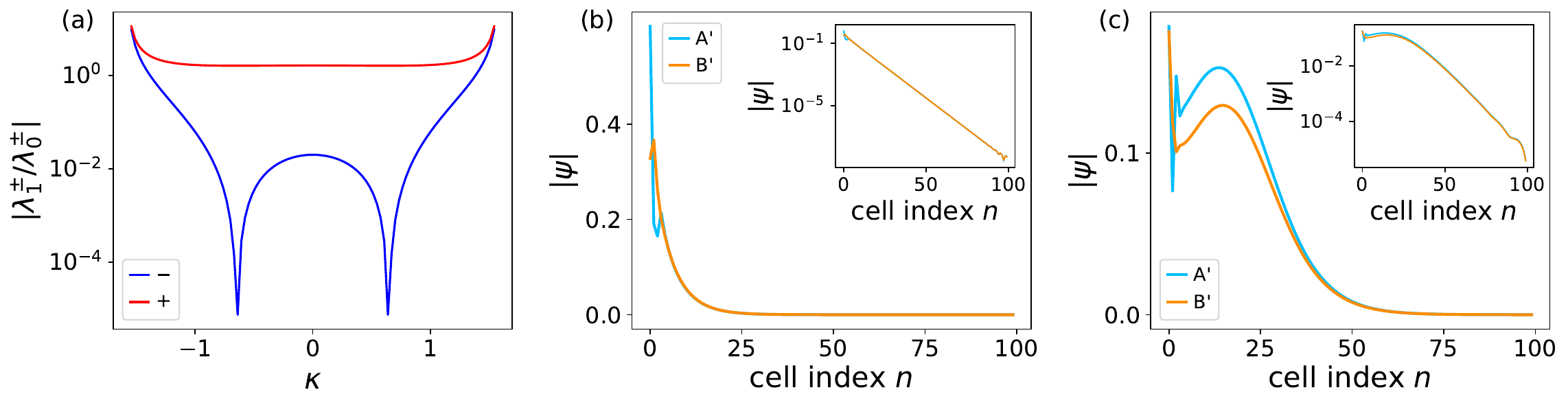}
\caption{\label{fig:fig2} (a) $|\lambda^{\pm}_1/\lambda^{\pm}_0|$ as a function of $\kappa$, under the condition that $\lim_{n\to\infty}{(\gamma_n - \gamma_{n-1})} = 0$. Here, $\kappa$ ranges from $-\pi/2$ to $\pi/2$. (b), (c) Profiles of eigenstates, with insets showing the eigenstates on a logarithmic scale. $\gamma_n = 20 \ln{(1+n/100)}$. For (b), the energy is $E = -0.270-0.079i$ ($\kappa = -0.560-0.185i$). For (c), $E = -0.485-0.025i$ ($\kappa = -1.272-0.168i$). Throughout (a)-(c), $t_1 = 0.4$, $t_2 = 0.5$.
}
\end{figure}

\begin{list}{}{\leftmargin=\parindent\rightmargin=0pt}
\item 
\textbf{(1.)} $\lim_{n\to\infty}{(\gamma_n - \gamma_{n-1})} = 0$. For example, consider logarithmic function $\gamma_n \sim \log n$, or a polynomial function of degree smaller than 1, i.e., $\gamma_n \sim n^\alpha$, where $0 < \alpha < 1$.

In this type of $\gamma_n$ function, it is apparent that the leading nontrivial term is the $1/\gamma_n$ term. By taking the first-order approximation of Eq.~(\ref{eqn:laurent}), we obtain:
\begin{equation}\label{eqn:laurent_2}
\begin{aligned}
    &b(n) = b_0 + \frac{b_1}{\gamma_n} + \cdots ,\ \ b_1 = \frac{2(t_1^2+t_2^2-E^2+2iEt_1)}{t_2}, \\
    &c(n) = c_0 + \frac{c_1}{\gamma_n} + \cdots ,\ \ c_1 = 4t_1.
\end{aligned}
\end{equation}
Similarly, the expansion of $\lambda^{\pm}(n)$ is 
\begin{equation}\label{eqn:lambda_expand}
    \lambda^{\pm}(n) = \lambda^{\pm}_0 + \frac{\lambda^{\pm}_1}{\gamma_n} + ...\ ,
\end{equation}
where $\lambda^{\pm}_1$ represents the expansion coefficient of the $1/\gamma_n$ term. By substituting Eqs.~(\ref{eqn:laurent_2}) and (\ref{eqn:lambda_expand}) into Eq.~(\ref{eqn:eigeneqn}), we can derive the ratio $\lambda^{\pm}_1/\lambda^{\pm}_0$:
\begin{equation}\label{eqn:lambda1/lambda0}
    \frac{\lambda^{
    \pm}_1}{\lambda^{\pm}_0} = -\frac{b_1 + c_1/\lambda^{\pm}_0}{2\lambda^{\pm}_0 + b_0}.
\end{equation}
The quantity $\lambda^{\pm}_1/\lambda^{\pm}_0$ characterizes the convergence speed of $ \lambda^{\pm}(n)$; specifically, a smaller value of $\lambda^{\pm}_1/\lambda^{\pm}_0$ indicates a smaller $1/\gamma_n$ term, resulting in a faster convergence to $\lambda_0^{\pm}$.

In the decoupled case of chains A and B, the `$-$' part of the T-shaped spectrum is characterized by $E^{0, -} = t_2 \sin k$, where $k = \left( \frac{i}{L+1} - \frac{1}{2}\right)\pi$, and $i = 1, 2, \ldots, L$. Analogously, we define $\kappa = \arcsin(E/t_2)$. Consequently, Eq.~(\ref{eqn:lambda_0}) becomes 
\begin{equation}\label{eqn:lambda_k}
    \lambda_0^{\pm} = \pm \exp(\pm i \kappa),    
\end{equation}
which corresponds to the Bloch phase factor $\pm e^{\pm ik}$. By substituting Eqs.~(\ref{eqn:laurent_2}) and (\ref{eqn:lambda_k}) into Eq.~(\ref{eqn:lambda1/lambda0}), we have
\begin{equation}\label{eqn:lambda1/lambda0_2}
    \frac{\lambda^{\pm}_1}{\lambda^{\pm}_0} = \mp \frac{(t_1 \pm t_2 \cos (\kappa))^2}{t_2 \cos (\kappa)}.
\end{equation}
In Section~\ref{section1} of the supplemental material, we argue that in the thermodynamic limit, the `$-$' part of the spectrum in the coupled case of chains A and B converges towards that of the decoupled cases. As a result, the parameter $\kappa$ in Eq.~(\ref{eqn:lambda1/lambda0_2}) can be closely approximated by $k$ in the decoupled case, which spans the range from $-\pi/2$ to $\pi/2$. Then,
it follows from Eq.~(\ref{eqn:lambda1/lambda0_2}) that $|\lambda^{-}_1/\lambda^{-}_0|$ is always smaller than $|\lambda^{+}_1/\lambda^{+}_0|$, as shown in Fig.~\ref{fig:fig2} (a). Especially, the ratio $|\lambda^{-}_1/\lambda^{-}_0|$ approaches zero when $\cos (\kappa)$ is close to $t_1/t_2$, which significantly accelerates the convergence of $\lambda^{-}(n)$ and exhibits typical ISSE [Fig.~\ref{fig:fig2} (b)]. As $\kappa$ approaches $\pm \pi/2$, the ratio $|\lambda^{\pm}_1/\lambda^{\pm}_0|$ diverges to infinity, which causes the ISSE to become less apparent [Fig.~\ref{fig:fig2} (c)].

\item
\textbf{(2.)} $\lim_{n\to\infty}{(\gamma_n - \gamma_{n-1})} = \gamma_0$, where $\gamma_0$ is a constant. For example, consider linear function $\gamma_n = \gamma_0 n$.

In this type of $\gamma_n$ function, both the $\frac{1}{\gamma_n}$ and $\frac{\gamma_n - \gamma_{n-1}}{\gamma_n}$ terms exhibit leading nontrivial contributions. Thus, the leading terms of $b(n)$, $c(n)$, and $\lambda^{\pm}(n)$ are given by
\begin{equation}
\begin{aligned}
    b(n) &= b_0 + \frac{b_1}{\gamma_n} + \cdots, \ \ b_1 = \frac{(t_1+iE)\gamma_0 + 2(t_1^2+t_2^2-E^2+2iEt_1) }{t_2}, \\
   c(n)  &= c_0 + \frac{c_1}{\gamma_n} + \cdots,  \ \ c_1 = 4t_1 + \gamma_0, \\
    \lambda^{\pm}(n) &= \lambda^{\pm}_0 + \frac{\lambda^{\pm}_1}{\gamma_n} + \cdots \ . 
\end{aligned}
\end{equation}
Since the forms of $b(n)$, $c(n)$, and $\lambda^{\pm}(n)$ are identical to those in \textbf{(1.)},  $\lambda^{\pm}_1/\lambda^{\pm}_0$ can be derived using the same method. This yields 
\begin{equation}\label{eqn:lambda1/lambda0_3}
    \frac{\lambda^{\pm}_1}{\lambda^{\pm}_0} = \mp \frac{(t_1 \pm t_2 \cos (\kappa)) (t_1 \pm t_2 \cos (\kappa) + \gamma_0/2)}{t_2 \cos(\kappa)}.
\end{equation}
The magnitude of the ratio, $|\lambda^{-}_1/\lambda^{-}_0|$, is generally smaller than  that of $|\lambda^{+}_1/\lambda^{+}_0|$. Notably, as $\cos (\kappa)$ approaches either $t_1/t_2$ or $(t_1 + \gamma_0/2)/t_2$, the value of $|\lambda^{-}_1/\lambda^{-}_0|$ tends to zero.

\item
\textbf{(3.)} $\lim_{n\to\infty}{(\gamma_n - \gamma_{n-1})} = \infty$, but $\lim_{n\to\infty}{\frac{\gamma_n - \gamma_{n-1}}{\gamma_n}} = 0$. For example, consider a polynomial function with a degree greater than $1$, $\gamma_n \sim n^\alpha$, where $\alpha > 1$.

In this case, the leading nontrivial term is $\frac{\gamma_n - \gamma_{n-1}}{\gamma_n}$. Upon expanding $b(n)$, $c(n)$, and $\lambda^{\pm}(n)$ to the first-order approximation, we obtain: 
\begin{equation}
\begin{aligned}
    b(n) & = b_0 + b_1\frac{\gamma_n - \gamma_{n-1}}{\gamma_n} +  \cdots, \ \ b_1 = \frac{t_1 + iE}{t_2}, \\
    c(n) & = c_0 + c_1\frac{\gamma_n - \gamma_{n-1}}{\gamma_n} +  \cdots, \ \ c_1 = 1, \\
    \lambda^{\pm}(n) & = \lambda^{\pm}_0 + \lambda^{\pm}_1\frac{\gamma_n - \gamma_{n-1}}{\gamma_n} + \cdots\ . 
\end{aligned}
\end{equation}
Using the same method as in \textbf{(1.)}, we can derive $\lambda^{\pm}_1/\lambda^{\pm}_0$, resulting in
\begin{equation}\label{eqn:lambda1/lambda0_4}
    \frac{\lambda^{\pm}_1}{\lambda^{\pm}_0} = \mp \frac{t_1 \pm t_2 \cos (\kappa)}{t_2 \cos (\kappa)}.
\end{equation}
Similar to the findings in \textbf{(1.)}, the magnitude of the ratio $|\lambda^{-}_1/\lambda^{-}_0|$ is always smaller than $|\lambda^{+}_1/\lambda^{+}_0|$. Additionally, this ratio converges to zero as $\cos (\kappa)$ approaches $t_1/t_2$.

\item
\textbf{(4.)} $\lim_{n\to\infty}{(\gamma_n - \gamma_{n-1})} = \infty$, and $\lim_{n\to\infty}{\frac{\gamma_n - \gamma_{n-1}}{\gamma_n}} = \mathrm{constant}$. For example, consider the exponential function $\gamma_n \sim \exp(\alpha n)$.

In this case, the term $\frac{\gamma_n - \gamma_{n-1}}{\gamma_n}$ degenerates to a constant in the expansion. Therefore, the leading nontrivial term is the $\frac{1}{\gamma_n}$ term. The coefficients $b(n)$ and $c(n)$ can be expanded as in Eq.~(\ref{eqn:laurent_2}),  with the same expressions of $b_1$ and $c_1$. However, the values of $b_0$ and $c_0$ will undergo a shift due to the term $\frac{\gamma_n - \gamma_{n-1}}{\gamma_n}$. This shift only affects the value of $\lambda^{\pm}_0$, while preserving the convergence properties of $\lambda^{\pm}(n)$.

\end{list}

\section{Recurrence relation of $\psi_n^{\pm}$ and its approximation}
In this section, we derive the recurrence relation for $\psi_n^{\pm}$ and discuss the approximations used in the main text.

We begin by decomposing $\ket{\psi(n)}$ into the eigenvectors of $T(n)$, using $\psi_n^{\pm} = \braket{\lambda^{\pm}_L(n) | \psi(n)}$ to denote the $\bra{\lambda^{\pm}_L(n)}$ component of $\ket{\psi(n)}$. This decomposition can be expressed as
\begin{equation}\label{eqn:decompose0}
   \ket{\psi(n)} = \psi_n^{+} \ket{\lambda^{+}_R(n)} + \psi_n^{-} \ket{\lambda^{-}_R(n)}.
\end{equation}
Substituting Eq.~(\ref{eqn:decompose0}) into Eq. (\ref{eqn:tm}) yields the recurrence relation between $\psi_{n-1}^{\pm}$ and $\psi_n^{\pm}$: 
\begin{equation}\label{eqn:recurrence0}
    \psi_n^{\pm}=\lambda^{\pm}(n)\Big[ \braket{\lambda^{\pm}_L(n)|\lambda^{\pm}_R(n-1)} \psi_{n-1}^{\pm} +\braket{\lambda^{\pm}_L(n)|\lambda^{\mp}_R(n-1)} \psi_{n-1}^{\mp}\Big].
\end{equation}
If $\gamma_n$ is uniform, i.e., the system has discrete translational symmetry, and the recurrence relation in Eq.~(\ref{eqn:recurrence0}) simplifies to $\psi_n^{\pm} = \lambda^{\pm}\psi_{n-1}^{\pm}$. This  result is straightforward, as the $\bra{\lambda^{\pm}_L(n)}$ component is scaled by the corresponding eigenvalue $\lambda^{\pm}$.
In more general scenarios, however, $\psi_n^{\pm} \ne \lambda^{\pm}(n)\psi_{n-1}^{\pm}$, which implies that if $\ket{\psi(n-1)}$ contains only the $\ket{\lambda^{+}_R(n-1)}$ component, then $\ket{\psi(n)}$ will have both the $\ket{\lambda^{+}_R(n)}$ and $\ket{\lambda^{-}_R(n)}$ components. This occurs because the transfer-matrix $T(n-1)$ differs from $T(n)$, introducing a mixed term $\braket{\lambda^{\pm}_L(n)|\lambda^{\mp}_R(n-1)} \psi_{n-1}^{\mp}$. However, if we assume that the transfer-matrix $T(n)$ varies slowly, then $\braket{\lambda^{\pm}_L(n)|\lambda^{\pm}_R(n-1)} \approx 1$ and $\braket{\lambda^{\pm}_L(n)|\lambda^{\mp}_R(n-1)} \approx 0$. Consequently, the recurrence relation in Eq.~(\ref{eqn:recurrence0}) can be approximated as follows:
\begin{equation}\label{eqn:coef}
    \psi_n^{+} \approx \lambda^{+}(n) \psi_{n-1}^{+},\ \ \ 
    \text{or} \ \ \ \psi_n^{-} \approx \lambda^{-}(n) \psi_{n-1}^{-},
\end{equation}
unless $|\psi_n^{-}| \gg |\psi_n^{+}|$ or  $|\psi_n^{+}| \gg |\psi_n^{-}|$, respectively.

\begin{figure}[b]
\includegraphics[width=0.9\textwidth]{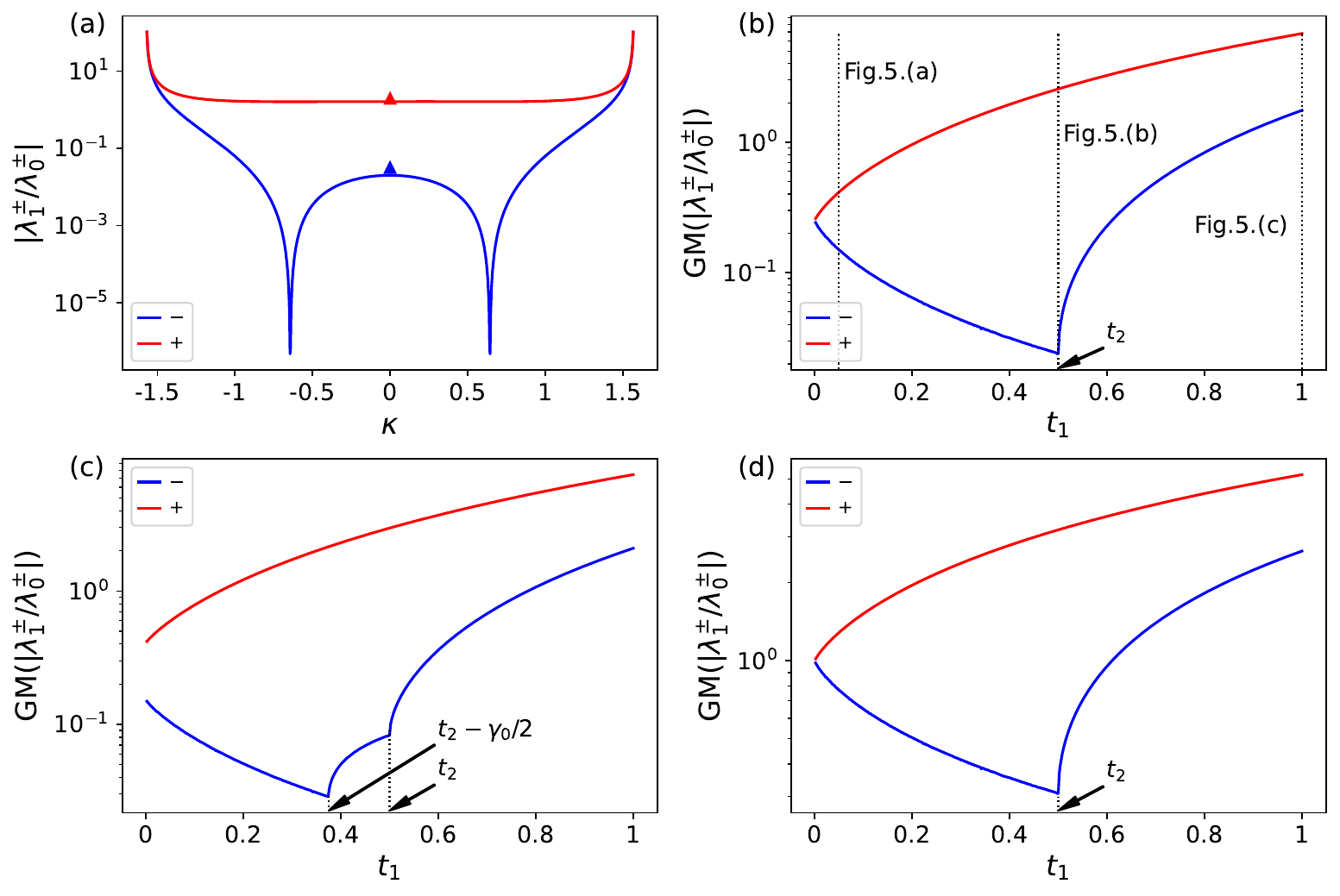}
\caption{\label{fig:convergence_speed} (a) $|\lambda^{\pm}_1/\lambda^{\pm}_0|$ as a function of $\kappa$ when $t_1$ and $t_2$ are fixed. Here, $\kappa$ ranges from $-\pi/2$ to $\pi/2$. The triangle represents the   geometric mean of $|\lambda^{\pm}_1/\lambda^{\pm}_0|$. The parameters are $t_1 = 0.4$, $t_2 = 0.5$, and $\gamma_n = 0.25n$. (b), (c), (d) Geometric mean of $|\lambda^{\pm}_1/\lambda^{\pm}_0|$ as a function of $t_1$, with $t_2 = 0.5$ fixed. The loss rate function $\gamma_n$ satisfies: $\lim_{n\to\infty}{(\gamma_n - \gamma_{n-1})} = 0$ for (b); $\lim_{n\to\infty}{(\gamma_n - \gamma_{n-1})} = \gamma_0 = 0.25$ for (c); and $\lim_{n\to\infty}{(\gamma_n - \gamma_{n-1})} = \infty$, while $\lim_{n\to\infty}{\frac{\gamma_n - \gamma_{n-1}}{\gamma_n}} = 0$ for (d). The three vertical dotted lines in (b) correspond to the parameter values of Fig.~\ref{fig:fig_4}.
}
\end{figure}

\section{ Analysis of parameters }

In this section, we discuss the impact of various parameters, including the hopping parameters $t_1$, $t_2$, and the chain length $L$, on the ISSE in our model.

\subsection{Hopping parameters $t_1$ and $t_2$}
In the main text, we state that to generate an evident ISSE, $\lambda^{-}(n)$ should converge to $\lambda^{-}_0$ rapidly and at a rate significantly faster than $\lambda^{+}(n)$, namely
\begin{equation}\label{eqn:converge_requirement}
\left| \lambda^{-}_1/\lambda^{-}_0 \right| \ll 1, \ \ \ \ \ \ \ \ \ \ \ \ \ \ \ \ \ \ \ \ \ \ \left| \lambda^{-}_1/ \lambda^{-}_0 \right| \ll \left| \lambda^{+}_1/\lambda^{+}_0 \right|.
\end{equation}
Subsequently, we investigate how the hopping parameters, $t_1$ and $t_2$, impact $\lambda^{\pm}_1/\lambda^{\pm}_0$ to satisfy this condition.

In Sec~\ref{sec3} of the supplemental material, we show that $\lambda^{\pm}_1/\lambda^{\pm}_0$ are determined by the parameters $t_1$, $t_2$, and $\kappa$. 
We also argued that in the thermodynamic limit, the parameter $\kappa$ can be effectively approximated by the corresponding $k$ in the decoupled case of chains A and B, with $k\in\left(  -\frac{\pi}{2},\frac{\pi}{2}\right) $. To characterize the convergence speed of $\lambda^{\pm}$ with fixed parameters $t_1$ and $t_2$, while excluding the influence of $\kappa$, we focus on the geometric mean \cite{Triola2012Statistics} of $|\lambda^{\pm}_1/\lambda^{\pm}_0|$ across $\kappa$ as a metric [Fig.~\ref{fig:convergence_speed}(a)]. This choice is motivated by the fact that the arithmetic mean of $|\lambda^{\pm}_1/\lambda^{\pm}_0|$ across $\kappa$ diverges in this context. The geometric mean (GM) is defined as
\begin{equation}\label{eqn:geometric_mean}
\mathrm{GM}\left(\left|\frac{\lambda^{\pm}_1}{\lambda^{\pm}_0}\right|\right) = 
\exp\left[\frac{1}{\pi} \int_{-\pi/2}^{\pi/2}{\ln\left( \left|\frac{\lambda^{\pm}_1}{\lambda^{\pm}_0}\right| \right) d\kappa}\right].
\end{equation}
We can depict $\mathrm{GM}(|\lambda^{\pm}_1/\lambda^{\pm}_0|)$ as a function of $t_1$ for fixed $t_2$ [Figs.~\ref{fig:convergence_speed}(b)-(d)]. 

\begin{figure}[t]
\includegraphics[width=1.0\textwidth]{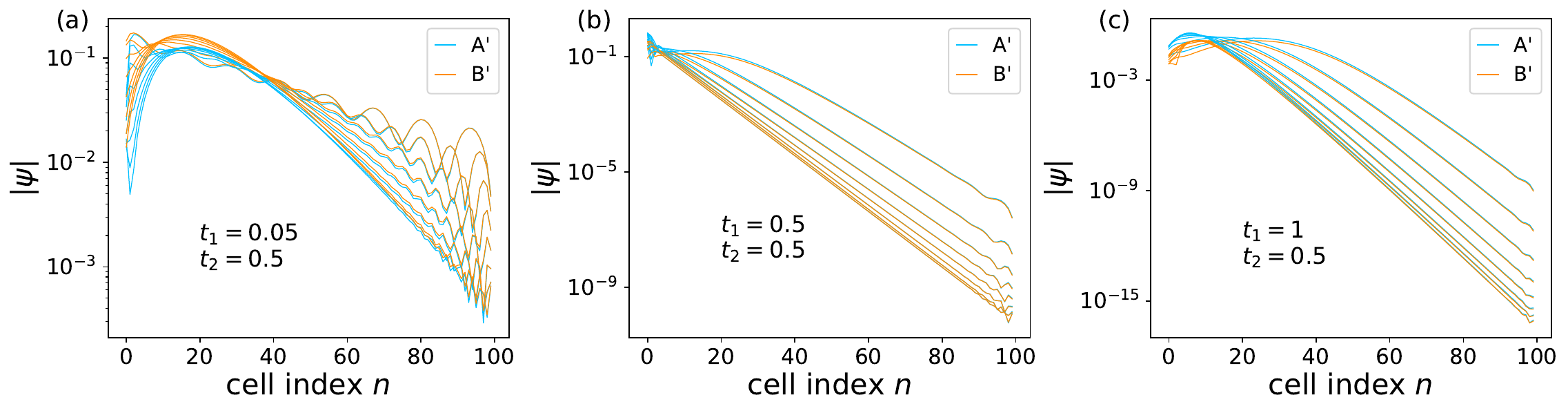}
\caption{\label{fig:fig_4} Profile of eight randomly chosen eigenstates from the `$-$' part of the spectrum. The parameters are $t_2 = 0.5$ and $\gamma_n = 20 \ln(1+n/100)$. For (a), $t_1 = 0.05$; for (b), $t_1 = 0.5$; and for (c), $t_1 = 1$. These parameter selections correspond to the illustration provided in Fig.~\ref{fig:convergence_speed}(b).
}
\end{figure}

As an illustrative example, we focus on the case where $\lim_{n\to\infty}{(\gamma_n - \gamma_{n-1})} = 0$  [Fig.~\ref{fig:convergence_speed}(b)]. A similar result can be shown for other types of the $\gamma_n$ function. Recall the convergence speed given by Eq.~(\ref{eqn:lambda1/lambda0_2}),
\begin{equation}\label{eqn:lambda1/lambda0_5}
    \frac{\lambda^{\pm}_1}{\lambda^{\pm}_0}(\kappa) = \mp \frac{(t_1 \pm t_2 \cos\kappa)^2}{t_2 \cos\kappa}.
\end{equation}
The critical point of Eq.~(\ref{eqn:lambda1/lambda0_5}) occurs at $t_1 = t_2$, since a real zero point $\kappa = \arccos(t_1/t_2)$ exists only when $t_1 \le t_2$. It  is evident  that $\mathrm{GM}(|\lambda^{+}_1/\lambda^{+}_0|)$ increases with $t_1$. In contrast,
 $\mathrm{GM}(|\lambda^{-}_1/\lambda^{-}_0|)$ decreases as $t_1$ increases from $0$, reaches a minimum at the critical point $t_1 = t_2$, and then increases with $t_1$. Therefore, when $t_1$ is close to
$t_2$, the condition stated in Eq.~(\ref{eqn:converge_requirement}) can be effectively satisfied, leading to typical ISSE behavior [Fig.~\ref{fig:fig_4}(b)]. However, if $t_1$ is significantly smaller or larger than $t_2$, the features of ISSE weaken, manifesting as an initial increase followed by a decrease in the wave function from left to right, resulting in a peak  near the left boundary [Fig.~\ref{fig:fig_4}(a), (c)].

\subsection{Chain length $L$}

In previous sections, we analyzed the model in the thermodynamic limit,  where the loss rate $\gamma_n$ exhibits a monotonic increase, ultimately diverging to infinity as $n\to \infty $. However, our analysis indicates that the emergence of an apparent ISSE in the model does not require the loss rate to reach this asymptotic condition.  Specifically, as long as the loss rate in the lattice is sufficiently increased so that $\lambda^{-}(n)$ approaches $\lambda^{-}_0$---that is, $\left|(\lambda^{-}(L) - \lambda^{-}_0)/\lambda^{-}_0\right| \ll 1$, where $L$ is the length of the lattice---we can expect to observe an apparent ISSE. This observation suggests that the ISSE can manifest even in finite-size systems, making it feasible for future experimental investigations.

\begin{figure}[t]
\includegraphics[width=0.8\textwidth]{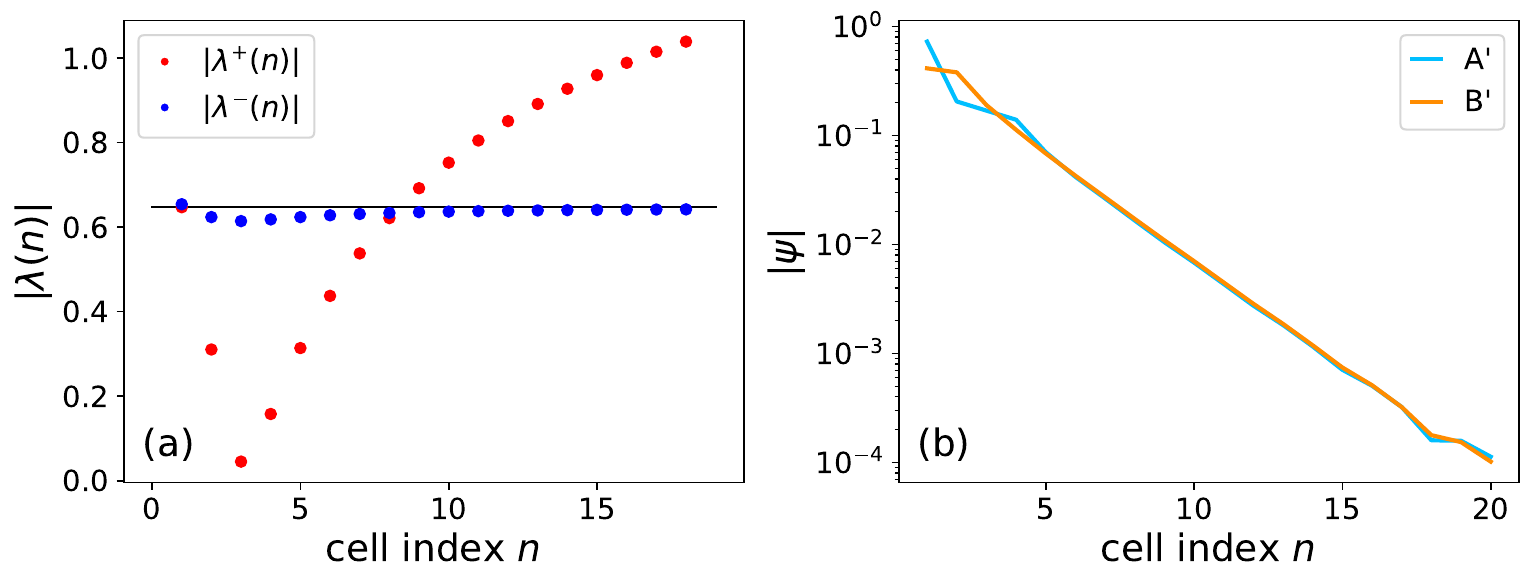}
\caption{\label{fig:fig_5} (a) Modulus of $\lambda^{\pm}(n)$ and (b) profile of eigenstate $|\psi|$ as functions of cell index $n$, for an eigenstate where $|\lambda^{-}_1/\lambda^{-}_0| \approx 3.78 \times 10^{-2}$, in a short lattice of length $L = 20$. The energy of this eigenstate is $E \approx -0.291-0.190i$.  The parameters are $t_1 = 0.4$, $t_2 = 0.5$, and $\gamma_n = 0.25 n$. The horizontal black line in (a) represent $\lambda^{-}_0$.
}
\end{figure}

To illustrate this phenomenon, we present a numerical example featuring a linearly increasing loss rate within a short lattice configuration. In this example, the lattice length $L$ is only $20$, and the loss rate at the right boundary, $\gamma_L$, is set to $5$, a value significantly below ``infinity". Nevertheless, this is adequate for $\lambda^{-}(n)$ to converge to $\lambda^{-}_0$ [Fig.~\ref{fig:fig_5}(a)], which is supported by the following calculation: 
\begin{equation}
    \left|\frac{\lambda^{-}(L) - \lambda^{-}_0}{\lambda^{-}_0}\right| \approx \left| \frac{\lambda^{-}_1}{\lambda^{-}_0} \frac{1}{\gamma_L} \right| \approx 7.2 \times 10^{-3} \ll 1.
\end{equation}
Consequently, an apparent ISSE is observed [Fig.~\ref{fig:fig_5}(b)], characterized by a uniform decay rate and the absence of interference. 

\bibliography{Supplemental}% Produces the bibliography via BibTeX.